\documentclass[fleqn,usenatbib,linenumbers]{mnras}
\usepackage{newtxtext,newtxmath}
\usepackage[T1]{fontenc}

\DeclareRobustCommand{\VAN}[3]{#2}
\let\VANthebibliography\thebibliography
\def\thebibliography{\DeclareRobustCommand{\VAN}[3]{##3}\VANthebibliography}

\usepackage{graphicx}	
\usepackage{amsmath}	
\usepackage{multirow}

\newcommand{\edit}[1]{{#1}}
\newcommand{\bhat}[1]{\hat{#1}}

\title[Non-perturbative magnetoasteroseismology]{Asteroseismic g-mode period spacings in strongly magnetic rotating stars}

\author[Rui, Ong, and Mathis]{
Nicholas Z. Rui$^{1}$\thanks{E-mail: nrui@caltech.edu}, J. M. Joel Ong$^{2,3}$, and St\'ephane Mathis$^{4}$
\\
$^{1}$TAPIR, California Institute of Technology, Pasadena, CA 91125, USA\\
$^{2}$Institute for Astronomy, University of Hawai`i at Mānoa, 2680 Woodlawn Dr., Honolulu, HI 96822, USA\\
$^{3}$NASA Hubble Fellow\\
$^{4}$Université Paris-Saclay, Université Paris Cité, CEA, CNRS, AIM, Gif-sur-Yvette F-91191, France
}

\pubyear{2023}

\begin{document}
\label{firstpage}
\pagerange{\pageref{firstpage}--\pageref{lastpage}}
\maketitle

\begin{abstract}
Strong magnetic fields are expected to significantly modify the pulsation frequencies of waves propagating in the cores of red giants or in the radiative envelopes of intermediate- and high-mass main-sequence stars.
We calculate the g-mode frequencies of stars with magnetic dipole fields which are aligned with their rotational axes, treating both the Lorentz and Coriolis forces non-perturbatively.
We provide a compact asymptotic formula for the g-mode period spacing, and universally find that strong magnetism decreases this period spacing substantially more than is predicted by perturbation theory.
These results are validated with explicit numerical mode calculations for realistic stellar models.
The approach we present is highly versatile: once the eigenvalues $\lambda$ of a certain differential operator are precomputed as a function of the magnetogravity and rotational frequencies (in units of the mode frequency), the non-perturbative impact of the Coriolis and Lorentz forces is understood under a broad domain of validity, and is readily incorporated into asteroseismic modeling.

\end{abstract}





\section{Introduction}

Because stellar oscillations extend throughout the stars in which they propagate, they contain a wealth of information about stellar interiors.
Asteroseismology is thus a sensitive probe of interior structure \citep{gough1993initial,christensen2012stellar,bellinger2017model,mombarg2020asteroseismic,bellinger2021asteroseismic,mombarg2022predictions,buldgen2022inversions}, rotation \citep{beck2012fast,mosser2012spin,deheuvels2012seismic,deheuvels2014seismic,kurtz2014asteroseismic,van2016interior,aerts2017interior,papics2017signatures,deheuvels2020seismic,burssens2023calibration,mombarg2023first}, mixing \citep{constantino2015treatment,li2018modelling,pedersen2018shape,michielsen2019probing,lindsay2023near}, evolution \edit{\citep{miglio2008probing,mosser2014mixed}}, and binary interaction history \citep{rui2021asteroseismic,deheuvels2022seismic,li2022discovery,tayar2022spinning,hekker2023low}.

In recent years, there has been a large amount of progress in developing asteroseismology as a probe of strong \textit{internal} magnetic fields, particularly through their effects on the gravity (g) modes which propagate in radiative regions.
Such fields likely have important consequences for the transport of angular momentum within evolved stars \citep{mathis2012low2,fuller2019slowing,aerts2019angular}.
On the red giant branch, g modes propagate in the radiative core, which may possess strong magnetic fields left over from efficient core convective dynamos on the main sequence \citep{fuller2015asteroseismology,stello2016prevalence}.
In these cases, magnetism may have a significant effect on the frequency spectrum: by measuring these frequency patterns, \citet{li202230} strongly constrain both the rotational periods and field strengths ($\gtrsim\!30\,\mathrm{kG}$) as well as their geometries for a modest sample of red giants.
Even stronger magnetic fields $\gtrsim\!100\,\mathrm{kG}$ are commonly invoked to explain the observed suppression of dipole ($\ell=1$) and quadrupole ($\ell=2$) oscillation modes in red giants \citep[e.g.,][]{garcia2014study,stello2016suppression,stello2016prevalence}.
Specifically, mode suppression is expected to occur in the non-perturbative ``strong magnetogravity'' regime \citep{fuller2015asteroseismology,lecoanet2017conversion,rui2023gravity}, when a mode's frequency $\omega$ is sufficiently close to the critical frequency
\begin{equation} \label{magnetogravity}
    \omega \lesssim \omega_{\mathrm{crit}} \sim \omega_B \equiv \sqrt{Nv_{Ar}/r}\mathrm{.}
\end{equation}

In \autoref{magnetogravity}, $v_{Ar}=B_r/\sqrt{4\pi\rho}$ is the radial component of the Alfv\'en velocity, $r$ is the radial coordinate, and $N$ is the Brunt--V\"ais\"al\"a (buoyancy) frequency, given by
\begin{equation}
    N^2 = g\left(\gamma^{-1}\frac{\mathrm{d}\ln p_0}{\mathrm{d}r} - \frac{\mathrm{d}\ln\rho_0}{\mathrm{d}r}\right)\mathrm{,}
\end{equation}

\noindent where $\gamma$ is the adiabatic index.

Equivalently, mode suppression occurs at some frequency $\omega_{\mathrm{crit}}$ when the magnetic field is at least comparable to some critical field \citep{fuller2015asteroseismology}:
\begin{equation}
    B \gtrsim B_{\mathrm{crit}} \propto \sqrt{\rho}\omega_{\mathrm{crit}}^2r/N\mathrm{.}
\end{equation}

Complementarily, main-sequence pulsators of intermediate mass ($\gtrsim\!\!1.3M_\odot$) have radiative, rather than convective, envelopes, and their g modes extend to their surfaces where they can be observed directly \citep{aerts2021probing}.
Examining the slowly pulsating B-type (SPB) star HD 43317, \citet{lecoanet2022asteroseismic} demonstrate that observations of mode suppression in main sequence (MS) pulsators may place meaningful constraints on their internal magnetism.
The detection of g modes in the vicinity of magnetic suppression suggests MS pulsators as a separate platform for testing the effect of strong magnetic fields on propagating gravity waves.

In the absence of effects such as magnetic fields or rotation, successive g modes are evenly spaced in period by a constant g-mode period spacing $\delta P_g$, which can be estimated as

\begin{equation} \label{zerofielddpg}
    \delta P_g = \frac{2\pi^2}{\sqrt{\ell(\ell+1)}}\left(\int\frac{N}{r}\,\mathrm{d}r\right)^{-1}\mathrm{,}
\end{equation}

\noindent where the integral is over the part of the radial cavity within which g modes propagate ($\omega<N$).

However, both rotation and magnetism leave distinctive signatures on the period spacing, both by lifting the degeneracy between modes of different $m$ (by breaking the spherical symmetry of the system) and by introducing period dependence \citep{bouabid2013effects,van2016interior,dhouib2022detecting}.
The period spacing as a function of period $\delta P_g=\delta P_g(P)$ is therefore a valuable measure for rotational and magnetic effects \citep{van2020detecting,henneco2021effect}.
Characterizing this observable non-perturbatively is the primary focus of this work.

Our paper proceeds as follows.
Section \ref{problemstatement} presents the problem statement and motivates the asymptotic treatment of magnetism and rotation.
Section \ref{analytic} derives the differential operator which governs the horizontal structure of magnetic gravito-inertial modes.
In Section \ref{horizontaleig}, we numerically calculate this operator's eigenvalues, which enter directly into an asymptotic formula for the period spacing.
In Section \ref{fullradial}, we solve the radial oscillation problem directly, including both magnetism and rotation while partially relaxing the asymptotic assumption.
Section \ref{results} presents the results of such calculations for models of red giants, $\gamma$ Doradus ($\gamma$ Dor), and SPB pulsators.
Finally, Section \ref{conclude} concludes.
The reader seeking our observational predictions is guided to Equations \ref{asymptotic} and \ref{transformdpg} (for the asymptotic period spacing) and the discussion in Section \ref{results}.

\section{Problem statement} \label{problemstatement}

The effect of magnetism on the asteroseismic period spacing has been previously explored by various authors.
So far, this work has typically either restricted its attention to toroidal fields \citep[$B_r\!=\!B_\theta\!=\!0$;][]{rogers2010interaction,mathis2011low,macgregor2011reflection,dhouib2022detecting}, treated magnetism perturbatively \citep[$B\!\ll\!B_{\mathrm{crit}}$;][]{cantiello2016asteroseismic,prat2019period,prat2020period,prat2020period2,mathis2021probing,bugnet2021magnetic,bugnet2022magnetic,li202230}, or worked in the ray-tracing limit \citep[$k_r,k_h\!\!\gg\!\!1/r$;][]{loi2018effects,loi2020magneto,loi2020effect}.
However, in realistic situations, gravity waves couple most strongly to the radial component of the field $B_r$ (see Section \ref{TARM}), which may be very strong \citep[$B\!\!\sim\!\!B_{\mathrm{crit}}$;][]{fuller2015asteroseismology}.
Moreover, due to geometric cancellation, observable modes are typically of low degree ($k_h\!\sim\!1/r$).

The central goal of this work is to calculate the period spacing pattern in the simultaneous presence of rotation and an axisymmetric \textit{radial} magnetic field in a non-perturbative way.
The work proceeds under the ``traditional approximation of rotation and magnetism'' defined in Section \ref{TARM} (which restricts attention purely to the radial field).

\subsection{The traditional approximation of rotation and magnetism (TARM)} \label{TARM}

Pure, low-frequency gravity waves follow the dispersion relation
\begin{equation} \label{pureg}
    \omega = \pm\frac{k_h}{k}N
\end{equation}

\noindent when $N\!\gg\!\omega$.
Therefore, their wavenumbers are primarily radial, with their radial wavenumbers exceeding their horizontal wavenumbers by ratios
\begin{equation}
    \frac{k_r}{k_h} \simeq \frac{N}{\omega}\mathrm{.}
\end{equation}

In the presence of restoring forces other than buoyancy or pressure (e.g., Coriolis forces, magnetic tension), the dispersion relation will be modified from \autoref{pureg}.
However, for modes which still have g mode character (i.e., buoyancy is still a significant restorative force), $k_r/k_h$ will still be comparable to $N/\omega\gg1$.
Throughout, we restrict our attention to modes whose wavenumbers are primarily radial: this is a crucial assumption of our work.

This approximation underlies the standard analytic treatment of gravity waves in rotating stars.
The qualitative behavior of low-frequency gravito-inertial waves can be seen in the dispersion relation in the fully Jeffreys--Wentzel--Kramers--Brillouin (JWKB) limit:
\begin{equation} \label{wkbrot}
    \omega^2 - \frac{k_h^2}{k^2}N^2 - (\bhat{k}\cdot\vec{\Omega})^2=0\mathrm{,}
\end{equation}

\noindent see, e.g., \citet{bildsten1996ocean} and \citet{lee1997low}.
Because low-frequency g modes have primarily radial wavenumbers, $k\!\approx \!k_r\!\sim\!(N/\omega)k_h\!\gg\!k_h$, the radial part of the rotation vector $\vec{\Omega}$ dominates in \autoref{wkbrot}.
It is thus both convenient and accurate for many purposes to assume $\vec{\Omega}\approx\Omega\cos\theta\,\bhat{r}$, i.e., to neglect the horizontal part of $\vec{\Omega}$.
Given its usefulness, this assumption is appropriately known as the ``traditional approximation of rotation'' (TAR).

We emphasize that the TAR is only valid when $k_r\!\!\gg\!\!k_h$.
It is therefore applicable when $\omega\!\!\ll\!\!N$ and $2\Omega\!\!\ll\!\!N$, with the interpretation that stratification is the dominant restorative force in the radial direction (such that the Coriolis force is only important in the horizontal directions).
The utility of this approximation is that it allows the (buoyancy-driven) vertical dynamics to be decoupled from the (Coriolis-driven) horizontal dynamics.
Because of this useful feature, the TAR has also found extensive use in geophysics \citep[e.g.,][]{eckart1960variation,longuet1968eigenfunctions}.
However, if either of the hypotheses of the TAR above are not satisfied, the traditional approximation should be abandoned \citep{dintrans1999gravito,dintrans2000oscillations,gerkema2005near,ballot2010gravity,mathis2014impact}.

When assumed, the TAR implies that the pressure perturbation varies in the horizontal directions according to the Laplace tidal equation:
\begin{equation} \label{lto}
    \begin{split}
        0 = \lambda p'(\mu) + \frac{\mathrm{d}}{\mathrm{d}\mu}&\left(\frac{1-\mu^2}{1-q^2\mu^2}\frac{\mathrm{d}p'(\mu)}{\mathrm{d}\mu}\right) - \frac{m^2}{\left(1-\mu^2\right)\left(1-q^2\mu^2\right)}p'(\mu) \\
        &- \frac{mq\left(1+q^2\mu^2\right)}{\left(1-q^2\mu^2\right)^2}p'(\mu)\mathrm{.}
    \end{split}
\end{equation}

In the non-rotating limit ($q\!\rightarrow\!0$), the Laplace tidal equation approaches the usual generalized Legendre equation, for which $\lambda\!=\!\ell(\ell+1)$ and the eigenfunctions are associated Legendre polynomials.
Here, $\mu=\cos\theta$ is the colatitude and $q=2\Omega/\omega$ is the spin parameter.
When computing mode frequencies the TAR, the effect of rotation is thus simply to replace $\ell(\ell+1)$ with $\lambda$.

To handle the effect of a strong dipole magnetic field, \citet{rui2023gravity} borrow intuition from the TAR.
The full JWKB dispersion relation for magnetogravity waves is
\begin{equation} \label{wkbmag}
    \omega^2 - \frac{k_h^2}{k^2}N^2 - (\vec{k}\cdot\vec{v}_A)^2 = 0\mathrm{,}
\end{equation}

\noindent where $\vec{v}_A=\vec{B}_0/\sqrt{4\pi\rho_0}$ is the Alfv\'en velocity, e.g., \citet{unno1989nonradial} and \citet{fuller2015asteroseismology}.
Analogously with the rotational argument, we see that the radial part of $\vec{B}$ dominates, and it suffices for a dipole magnetic field to assume $\vec{B}\approx B_0\cos\theta\,\bhat{r}$.
The pressure perturbation then follows
\begin{equation} \label{mto}
    0 = \lambda p'(\mu) + \frac{\mathrm{d}}{\mathrm{d}\mu}\left(\frac{1-\mu^2}{1-b^2\mu^2}\frac{\mathrm{d}p'(\mu)}{\mathrm{d}\mu}\right) - \frac{m^2}{\left(1-\mu^2\right)\left(1-b^2\mu^2\right)}p'(\mu)\mathrm{,}
\end{equation}

\noindent where $b=k_rv_{Ar}/\omega$.
The interpretation of this approximation is that the fluid is sufficiently stratified that buoyancy is the only important restorative force in the radial direction (i.e., the Lorentz force need only be included in the horizontal directions).
As in the TAR, including magnetism in a calculation of mode frequencies under this approximation simply involves replacing $\ell(\ell+1)$ with a suitably computed $\lambda$ when solving the radial problem.

We note the similar forms of \autoref{lto} (for rotation) and \autoref{mto} (for magnetism).
However, unlike the singularities in \autoref{lto} (around which the eigenfunctions are smooth), the singularities in \autoref{mto} are of significantly different character, and imply sharp fluid features corresponding to resonances with Alfv\'en waves \citep{rui2023gravity}.
For the frequency-shift analysis conducted in this work, this property of the singularities in \autoref{mto} motivates restriction to solutions for which $b<1$ (so that the Alfv\'en resonances are not on the domain).

In this work, we generalize both the traditional approximation of rotation and its magnetic analogue to incorporate both effects: in other words, we consider only the effects of the \textit{radial} components of both the rotation vector and magnetic field.
Equivalently, we include only the horizontal components of the Coriolis and Lorentz forces.
Hereafter, we refer to this joint approximation as the traditional approximation of rotation and magnetism (TARM).

\subsection{Assumptions, conventions, and scope} \label{assumptions}

In addition to assuming that $k_r\!\gg\!k_h$, we adopt the JWKB approximation in the radial direction \textit{only}, i.e., we assume that the equilibrium structure and field of the star vary on length scales much larger than the radial wavelength (the ``asymptotic'' regime).
Because such length scales are typically $\sim r$, this assumption is usually justified, although it may be violated in the presence of sharp compositional gradients which are known to produce mode-trapping phenomena \citep[e.g.,][]{miglio2008probing,pedersen2018shape,michielsen2019probing}.
In Section \ref{fullradial}, we solve for the full radial dependence of the wavefunction without directly assuming that the radial wavenumber is large.
However, under the TARM, we perform this calculation using a precomputed grid of horizontal eigenvalues $\lambda$ (see Section \ref{horizontaleig}) which \textit{does} make this assumption.
Therefore, the calculation described in Section \ref{fullradial} is expected to partially, but not fully, capture non-JWKB effects in the radial direction.

We index branches by the angular degree $\ell$ and order $m$.
In particular, a mode is said to have some value of $\ell$ and $m$ when the horizontal dependence becomes the spherical harmonic $Y_{\ell m}$ when both the field and rotation are smoothly taken to zero.
We caution that, while we may refer to some mode as having some degree $\ell$ in a rotating and/or magnetized star, $Y_{\ell m}$ is not the correct horizontal dependence, and the eigenvalues are no longer $\ell(\ell+1)$.
For the angular order $m$, we adopt the sign convention used by \citet{lee1997low} and \citet{rui2023gravity} that $mq>0$ ($mq<0$) corresponds to retrograde (prograde) modes.
Additionally, without loss of generality, we consider throughout the case where $q\!>\!0$ and $b\!>0\!$ (which appear in, e.g., Equations \ref{lto} and \ref{mto}, respectively), i.e., positive (negative) azimuthal order $m$ corresponds to retrograde (prograde) modes.
In this problem, the sign of $b$ is irrelevant, and the effect of a sign change in $q$ can be fully compensated by changing the sign convention of $m$.

In the presence of (solid-body) rotation, it is important to distinguish the mode frequency in the inertial frame (which is observable) from the mode frequency in the frame co-rotating with the star (in which the effect of rotation appears as a Coriolis force).
Hereafter, we use $\omega$ ($\bar{\omega}$) to denote the mode frequency in the inertial (co-rotating) frame.
Hence, we calculate the oscillation modes directly with respect to $\bar{\omega}$, but convert to $\omega$ for observational purposes.

We restrict our attention to a magnetic field whose radial part has a dipolar horizontal dependence.
However, our results are not sensitive to the radial dependence of the field (as long as it is not very steep), or the geometry of the horizontal field components (as long as they are not much larger than the radial component).
While Section \ref{asymptotictext} makes no additional assumptions about the field than those listed above, Section \ref{fullradial} requires a radial magnetic field profile $B_{0r}=B_{0r}(r)$.
For this work, we adopt the Prendergast magnetic field geometry \citep{prendergast1956equilibrium}.
For our purposes, it suffices to specify the radial component of the magnetic field:
\begin{equation}
    B_{0r} = B_c\frac{2R^2}{r^2}\frac{\beta}{\Lambda^2}\left(\frac{r^2}{R^2} - \frac{r}{R}\frac{j_1(\Lambda r)}{j_1(\Lambda)}\right)\cos\theta\mathrm{,}
\end{equation}

\noindent where $j_1(x)=\sin x/x^2-\cos x/x$ is the first spherical Bessel function and $R$ is the radius of the star.
Although \citet{kaufman2022stability} have recently shown that the Prendergast geometry is likely unstable over timescales relevant to stellar evolution, we adopt it simply as a closed-form model for a large-scale, dipole-like field, and we expect our findings to be insensitive to the exact radial dependence of the field.
Following, e.g., \citet{kaufman2022stability}, we take $\Lambda\approx5.76346$ and $\beta\approx1.31765$, corresponding to the normalized, lowest-energy field solution with a vanishing surface field.
Hereafter, $B_c$ should be understood to refer to the \textit{radial} component of the core magnetic field amplitude, although it is typically expected that the radial and horizontal components of the field are comparable.
We expect all of the chief results of this work to be robust to magnetic field geometry, as long as $B_{0h}/B_{0r}\lesssim N/\omega$ and $k_r\gg\mathrm{d}\ln B_{0r}/\mathrm{d}r$.

We specialize to the case where magnetism is not strong enough to suppress the modes (although we explore the mode frequencies right up to this limit).
While the suppression mechanism of magnetogravity waves is not fully understood, suppression may occur when magnetogravity waves refract upwards at some critical $\omega\!=\!\omega_B$ to infinite wavenumber \citep{lecoanet2017conversion,lecoanet2022asteroseismic,rui2023gravity} or are damped out by phase-mixing processes once resonant with Alfv\'en waves ($b>1$) in a manner similar to that described by \citet{loi2017torsional}.
Therefore, we restrict the scope of our calculations to the case where $b\!=\!k_rv_{Ar}/\omega\!<\!1$ and $\omega\!<\!\omega_B$.
Under these circumstances, the effects of magnetism on g modes should be well-modeled by our method.


For demonstrative purposes, we restrict most of our attention in this work to the dipole ($\ell\!=\!1$) and quadrupole ($\ell\!=\!2$) modes, although our calculations do not assume this, and it is not more complicated to extend this analysis to higher $\ell$.
Low-degree g modes suffer the least from geometric cancellation and are thus the easiest g modes to observe (there are no radial g modes).
For simplicity, we assume modes are adiabatic, and neglect perturbations to the gravitational potential (i.e., we adopt the Cowling approximation).
The general result that the perturbative theory underestimates the impact of magnetism on the period spacings for the dipole modes (Section \ref{results}) is also expected to hold for the quadrupole modes, although the asymmetry in the frequency shifts of different multiplets is known to behave differently \citep[cf. Section \ref{redgiant} of][]{bugnet2021magnetic}.

\section{Analytic formulation} \label{analytic}

In this Section, we derive an expression for the horizontal equation obeyed by low-frequency g modes under the simultaneous influence of uniform (or weak differential) rotation and a dipolar magnetic field (Section \ref{gravity}).
Under the TARM, the eigenvalues associated with these normal modes can be easily translated to an asymptotic expression for the period spacing (Section \ref{asymptotictext}).

\subsection{Fluid equations for gravity modes} \label{gravity}

In the presence of gravity, magnetic tension and pressure, and Coriolis forces, the linearized momentum equation is
\begin{equation} \label{momentum}
    \begin{split}
        \rho_0\partial_t^2\vec{\xi} + 2\rho_0&\Vec{\Omega}\times\partial_t\vec{\xi} \\
        &= -\nabla\left(p' + \frac{1}{4\pi}\vec{B}_0\cdot\vec{B}'\right) - \rho'g\bhat{r} + \frac{1}{4\pi}\left(\vec{B}_0\cdot\nabla\right)\vec{B}'
    \end{split}
\end{equation}

\noindent where $\Vec{\xi}$ is the fluid displacement, subscript $0$ and primes denote equilibrium and perturbed quantities respectively, $g$ is the gravitational acceleration, and
\begin{equation}
    \vec{B}' = \left(\vec{B}_0\cdot\nabla\right)\vec{\xi}
\end{equation}

\noindent is the magnetic field perturbation.
Equations \ref{momentum} ignore the centrifugal force, and apply a Cowling approximation to neglect perturbations in $g$.
Under the TARM, $\partial_r\rightarrow-ik_r$ when acting on a perturbation, and the magnetic tension term in \autoref{momentum} thus becomes
\begin{equation}
    \frac{1}{4\pi}\left(\vec{B}_0\cdot\nabla\right)\vec{B}' = -\frac{k_r^2B_0^2}{4\pi}\vec{\xi} \equiv \rho_0k_r^2v_{Ar}^2\cos^2\theta\vec{\xi}
\end{equation}

\noindent where $|v_{Ar}\cos\theta|$ is the radial component of the Alfv\'en velocity, with the angular dependence explicitly factored out.

In spherical coordinates and applying the traditional approximation, the momentum equation becomes
\begin{subequations} \label{momentum2}
    \begin{gather}
        -\rho_0\bar{\omega}^2\xi_r = ik_rp' - \rho'g - \rho_0k_r^2v_{Ar}^2\cos^2\theta\xi_r \label{radial_mtm} \\
        -\rho_0\bar{\omega}^2\xi_\theta - 2i\rho_0\bar{\omega}\Omega\cos\theta\,\xi_\phi = -\frac{1}{r}\frac{\mathrm{d}p'}{\mathrm{d}\theta} - \rho_0k_r^2v_{Ar}^2\cos^2\theta\xi_\theta \\
        -\rho_0\bar{\omega}^2\xi_\phi + 2i\rho_0\bar{\omega}\Omega\cos\theta\,\xi_\theta = -\frac{im}{r\sin\theta}p' - \rho_0k_r^2v_{Ar}^2\cos^2\theta\xi_\phi
    \end{gather}
\end{subequations}

\noindent where we have assumed harmonic time dependence, $\partial_t\rightarrow i\bar{\omega}$, and used axisymmetry to take $\partial_\phi\rightarrow im$.
Magnetic tension dominates over magnetic pressure in the asymptotic regime, and so the latter is ignored in Equations \ref{momentum2}.

For adiabatic oscillations, the pressure and density $p$ and $\rho$ are related by
\begin{equation} \label{adiabatic}
    \frac{\mathrm{D}\ln p}{\mathrm{D}t} = \gamma\frac{\mathrm{D}\ln \rho}{\mathrm{D}t}\mathrm{,}
\end{equation}

\noindent where $\mathrm{D}/\mathrm{D}t=\partial/\partial t+\Vec{u}\cdot\nabla$ denotes the advective derivative.
\autoref{adiabatic} can be linearized to
\begin{equation} \label{brunt1}
    \rho' = \rho_0N^2\xi_r/g + p'/c_s^2
\end{equation}

\noindent where $c_s=\sqrt{\gamma p_0/\rho_0}$ is the sound speed.
\edit{For gravity waves, the first term dominates, so that}
\begin{equation}
    \edit{\rho' \approx \rho_0N^2\xi_r/g.}
\end{equation}

Finally, the fluid perturbation must satisfy the equation of continuity, so that
\begin{equation} \label{continuity1}
    \edit{\nabla\cdot\Vec{\xi} = 0\mathrm{,}}
\end{equation}

\noindent \edit{where we have applied the Boussinesq approximation \citep{proctor1982magnetoconvection}.}

Now, the horizontal momentum equations give a linear system of equations for $\xi_\theta$ and $\xi_\phi$ in terms of $p'$:
\begin{subequations} \label{horizontal}
    \begin{gather}
        (1-b^2\cos^2\theta)\xi_\theta + iq\cos\theta\,\xi_\phi = \frac{1}{\rho_0\bar{\omega}^2r}\frac{\mathrm{d}p'}{\mathrm{d}\theta} \\
        -iq\cos\theta\,\xi_\theta + (1-b^2\cos^2\theta)\xi_\phi = \frac{im}{\rho_0\bar{\omega}^2r\sin\theta}p'
    \end{gather}
\end{subequations}

\noindent where $b\!=\!k_rv_{Ar}/\bar{\omega}$ \citep{rui2023gravity} and again $q\!=\!2\Omega/\bar{\omega}$ \citep{lee1997low} are the dimensionless parameters governing the effects of magnetism and rotation on the horizontal eigenfunctions.
Equations \ref{horizontal} can be solved to obtain
\begin{subequations} \label{horizontal_perts}
    \begin{gather}
        \xi_\theta = \frac{\sqrt{1-\mu^2}}{\rho_0\bar{\omega}^2r\left[(1-b^2\mu^2)^2 - q^2\mu^2\right]}\left(\frac{mq\mu}{1-\mu^2} p' - (1-b^2\mu^2)\frac{\mathrm{d}p'}{\mathrm{d}\mu}\right) \\
        \xi_\phi = \frac{i\sqrt{1-\mu^2}}{\rho_0\bar{\omega}^2r\left[(1-b^2\mu^2)^2 - q^2\mu^2\right]}\left((1-b^2\mu^2)\frac{m}{1-\mu^2}p' - q\mu\frac{\mathrm{d}p'}{\mathrm{d}\mu}\right)
    \end{gather}
\end{subequations}

\noindent where $\mu=\cos\theta$. Likewise, the radial component of the momentum equation (\autoref{radial_mtm}) can be \edit{solved to yield}
\begin{equation} \label{radial_pert}
    \edit{\xi_r = \frac{ik_r}{\rho_0N^2}p'\mathrm{.}}
\end{equation}

Substituting Equations \ref{brunt1}, \ref{horizontal_perts}, and \ref{radial_pert} into the continuity equation (\autoref{continuity1}), we obtain
\begin{equation} \label{operatorintermediate}
    \edit{\mathcal{L}^{m,b,q}[p'] + \frac{\bar{\omega}^2r^2k^2}{N^2}p' = 0\mathrm{,}}
\end{equation}

\noindent where the differential operator $\mathcal{L}^{m,b,q}$ is given by
\begin{equation} \label{L}
    \begin{split}
        \mathcal{L}&^{m,b,q}[f(\mu)] = \frac{\mathrm{d}}{\mathrm{d}\mu}\left(\frac{(1-\mu^2)(1-b^2\mu^2)}{(1-b^2\mu^2)^2-q^2\mu^2}\frac{\mathrm{d}f(\mu)}{\mathrm{d}\mu}\right) \\
        &- \frac{m^2}{1-\mu^2}\frac{1-b^2\mu^2}{(1-b^2\mu^2)^2-q^2\mu^2}f(\mu) \\
        &- mq\left(\frac{4b^2\mu^2(1-b^2\mu^2)+2q^2\mu^2}{\left[(1-b^2\mu^2)^2-q^2\mu^2\right]^2} + \frac{1}{(1-b^2\mu^2)^2-q^2\mu^2}\right)f(\mu)\mathrm{.}
    \end{split}
\end{equation}


The operator $\mathcal{L}^{m,b,q}$ further reduces to the standard Laplace tidal operator \citep[e.g.,][]{lee1997low} when $b=0$ (no magnetism), and to the magnetic operator discussed by \citet{rui2023gravity} when $q=0$ (no rotation). Hereafter, we define the ``eigenvalues'' $\lambda$ of $\mathcal{L}^{m,b,q}$ as constants admitting solutions $f(\mu)$ to
\begin{equation} \label{eig}
    \mathcal{L}^{m,b,q}[f(\mu)] + \lambda f(\mu) = 0\mathrm{,}
\end{equation}

\noindent i.e., the eigenvalues of $\mathcal{L}^{m,b,q}$ in the ``standard'' sign convention are $-\lambda$.

When $b\!=\!q\!=\!0$ (i.e., no magnetism or rotation), $\mathcal{L}^{m,b,q}[f(\mu)]$ reduces further still to the standard Laplacian operator on a sphere, where solutions to the associated boundary value problem are the spherical harmonics, indexed by integers $\ell, m$ with eigenvalues $\ell(\ell + 1)$. \edit{\autoref{operatorintermediate}} becomes
\begin{equation} \label{magnetogravitydisp}
    \bar{\omega}^2 = \frac{\lambda/r^2}{k_r^2}N^2 = \frac{\Tilde{k}_h^2}{k_r^2}N^2\mathrm{,}
\end{equation}

\noindent where $\Tilde{k}_h\equiv\sqrt{\lambda}/r$ is an effective horizontal wavenumber, which incorporates the effects of rotation and magnetism. By analogy with the spherically symmetric case, we may define an effective degree
\begin{equation} \label{effective}
    \ell_e = \sqrt{\lambda + 1/4} - 1/2\mathrm{,}
\end{equation}

\noindent such that $\lambda=\ell_e(\ell_e+1)$.
In the TARM, oscillation modes are calculated by replacing $\ell$ with $\ell_e$ throughout the entire star, in the same manner as is done in the standard TAR.


\subsection{Asymptotic period spacing} \label{asymptotictext}

In the absence of rotation and magnetism, gravity modes obey the dispersion relation $\bar{\omega}=\pm k_hN/k_r\propto k_r^{-1}$.
In the asymptotic regime (where $k_rr\rightarrow\infty$), this implies that adjacent g modes (with relative radial orders $\delta n_g=1$, and $k_r\sim n_g/r$) are spaced uniformly in the mode period $P$.
In this Section, we derive an expression for the asymptotic period spacing $\delta P_g$ for g modes.
We note that further departures from the asymptotic formula are expected when the stellar structure varies over a comparable radial scale to the wavefunction, or when there is mode mixing.

Before proceeding, we review a fundamental difference between the inclusion of uniform rotation and magnetism.
For rotation, the fluid equations are solved by eigenfunctions whose shapes are solely parameterized by the spin parameter $q=2\Omega/\bar{\omega}$, which can be calculated using stellar model parameters and the mode frequency, i.e., without knowledge of $k_r$.
Observed spin parameters for intermediate-mass g-mode pulsations range from $q\simeq0.1$ to $q\simeq30$ \citep{aerts2017interior}.
However, for magnetism, the parameter which controls the shapes of the eigenfunctions, $b=k_rv_{Ar}/\bar{\omega}$, \textit{does} depend on $k_r$ (which varies mode-to-mode and with $r$ in a complicated way).
Fortunately, \autoref{magnetogravitydisp} can also be rewritten
\begin{equation} \label{lambdab2a2}
    \lambda = b^2/a^2\mathrm{,}
\end{equation}

\noindent where the parameter $a$ \citep[described by][]{rui2023gravity} is given by
\begin{equation}
    a = \frac{Nv_{Ar}}{r\bar{\omega}^2}\mathrm{.}
\end{equation}

This parameter is the squared ratio of the magnetogravity frequency $\omega_B$ (\autoref{magnetogravity}) to the mode frequency ($a\sim\omega_B^2/\bar{\omega}^2$) and, conveniently, \textit{can} be computed in terms of the stellar model and $\bar{\omega}$ alone.
By computing the horizontal eigenfunctions $\lambda$ as a function of $b$ and then inverting \autoref{lambdab2a2}, $\lambda$ can be found as a function of $a$.

To compute the period spacing in the co-rotating frame, we first observe that the radial phase $\varphi_g$ across the gravity mode cavity is
\begin{equation} \label{phase}
    \varphi_g = \pi(n_g + \epsilon_g) = \int k_r\mathrm{d}r = \frac{\bar{P}}{2\pi}\int\sqrt{\lambda}\,\frac{N}{r}\,\mathrm{d}r\mathrm{,}
\end{equation}

\noindent where we have used \autoref{magnetogravitydisp}, and the integral is over the region of the star where $\omega\!<\!N$ and $\omega\!<\!k_hc_s$.
In \autoref{phase}, $n_g$ is the radial order, and $\epsilon_g$ is a (here unimportant) phase offset.
Adjacent modes (with $\delta n_g=1$) will thus have
\begin{equation} \label{radialorder}
    \pi\delta n_g = \pi = \frac{\delta\bar{P}_g}{2\pi}\int\sqrt{\lambda}\,\frac{N}{r}\,\mathrm{d}r + \frac{\pi\bar{P}}{4\pi}\int\frac{\delta\lambda}{\sqrt{\lambda}}\frac{N}{r}\,\mathrm{d}r\mathrm{,}
\end{equation}

\noindent where we have neglected the frequency dependence of the bounds of the buoyancy integral.

Because $q\propto\bar{\omega}^{-1}\propto\bar{P}$ and $a\propto\bar{\omega}^{-2}\propto\bar{P}^2$,
\begin{equation} \label{lambdaderiv}
    \delta\lambda = \frac{\mathrm{d}\lambda}{\mathrm{d}\bar{P}}\delta\bar{P}_g = \frac{\lambda}{\bar{P}}\frac{\mathrm{d}\ln\lambda}{\mathrm{d}\ln\bar{P}}\delta\bar{P}_g = \frac{\lambda}{\bar{P}}\left(\frac{1}{2}\frac{\mathrm{\partial}\ln\lambda}{\mathrm{\partial}\ln q} + \frac{\mathrm{\partial}\ln\lambda}{\mathrm{\partial}\ln a}\right)\delta\bar{P}_g\mathrm{.}
\end{equation}

Combining Equations \ref{radialorder} and \ref{lambdaderiv} and solving for $\delta\bar{P}_g$ gives
\begin{equation} \label{asymptotic}
    \delta\bar{P}_g = 2\pi^2\left(\int\sqrt{\lambda}\left(1 + \frac{1}{2}\frac{\mathrm{\partial}\ln\lambda}{\mathrm{\partial}\ln q} + \frac{\mathrm{\partial}\ln\lambda}{\mathrm{\partial}\ln a}\right)\frac{N}{r}\,\mathrm{d}r\right)^{-1}\mathrm{.}
\end{equation}

This approaches the well-known, zero-field, zero-rotation asymptotic formula in the relevant limit (\autoref{zerofielddpg}), as well as Equation 4 of \citet{bouabid2013effects} which was derived for the purely rotational case.

\autoref{asymptotic} requires the calculation of $\left(\partial\ln\lambda/\partial\ln q\right)_a$ and $\left(\partial\ln\lambda/\partial\ln a\right)_q$, where subscripts denote fixed variables with respect to the partial derivative.
In Section \ref{collocation}, we compute $\lambda$ and its derivatives on a discrete, rectangular grid of $b$ and $q$.
While $\left(\partial\ln\lambda/\partial\ln a\right)_q$ is easy to calculate numerically via a finite difference formula (since fixing $q$ is straightforward), computing $\left(\partial\ln\lambda/\partial\ln q\right)_a$ is slightly trickier because it is harder to fix $a$.
Via \autoref{lambdab2a2}, we see that
\begin{equation}
    \left(\frac{\partial\lambda}{\partial q}\right)_a = \frac{2b}{a^2}\left(\frac{\partial b}{\partial q}\right)_a\mathrm{.}
\end{equation}

Using the identity that
\begin{equation}
    -1 = \left(\frac{\partial b}{\partial q}\right)_a\left(\frac{\partial a}{\partial b}\right)_q\left(\frac{\partial q}{\partial a}\right)_b\mathrm{,}
\end{equation}

\noindent we obtain
\begin{equation}
    \left(\frac{\partial\lambda}{\partial q}\right)_a = -\frac{2b}{a^2}\left(\frac{\partial a}{\partial q}\right)_b\left(\frac{\partial a}{\partial b}\right)_q^{-1}
\end{equation}

\noindent so that
\begin{equation} \label{magic}
    \left(\frac{\partial\ln\lambda}{\partial\ln q}\right)_a = -\frac{2q}{b}\left(\frac{\partial a}{\partial q}\right)_b\left(\frac{\partial a}{\partial b}\right)_q^{-1}\mathrm{.}
\end{equation}
We use \autoref{magic} in our numerical calculation of $\delta P_g$.

In the \textit{inertial} frame, the observed frequencies $\omega$ are related to $\bar{\omega}$ under our sign convention by
\begin{equation} \label{change}
    \omega = \bar{\omega} - m\Omega\mathrm{,}
\end{equation}

\noindent so that the periods $P$ and $\bar{P}$ in the inertial and co-rotating frames are related by
\begin{equation} \label{converttoinertial}
    P = \frac{\bar{P}}{1 - m\bar{P}/P_{\mathrm{rot}}}\mathrm{,}
\end{equation}

\noindent where $P_{\mathrm{rot}}$ is the rotation period.
The (asymptotic) period spacing measured by an observer is thus given by
\begin{equation} \label{transformdpg}
    \delta P_g = \frac{\delta\bar{P}_g}{(1-m\bar{P}/P_{\mathrm{rot}})^2}\mathrm{.}
\end{equation}

Thus, the inclusion of either rotation or magnetism will also leave distinct imprints on $\delta P_g$ as a function of mode period: understanding these signatures is crucial for extracting these properties from $\delta P_g$.


\section{Numerical solutions of the horizontal problem} \label{horizontaleig}

In preceding sections, we have introduced an analytic formulation of the magnetorotational pulsation problem.
However, applying the TARM to concrete predictions of oscillation spectra requires robust numerical solutions for the horizontal eigenvalues $\lambda$.
We describe our numerical procedure for this calculation in this Section.

\subsection{Numerical collocation scheme} \label{collocation}

\begin{figure*}
    \centering
    \includegraphics[width=\textwidth]{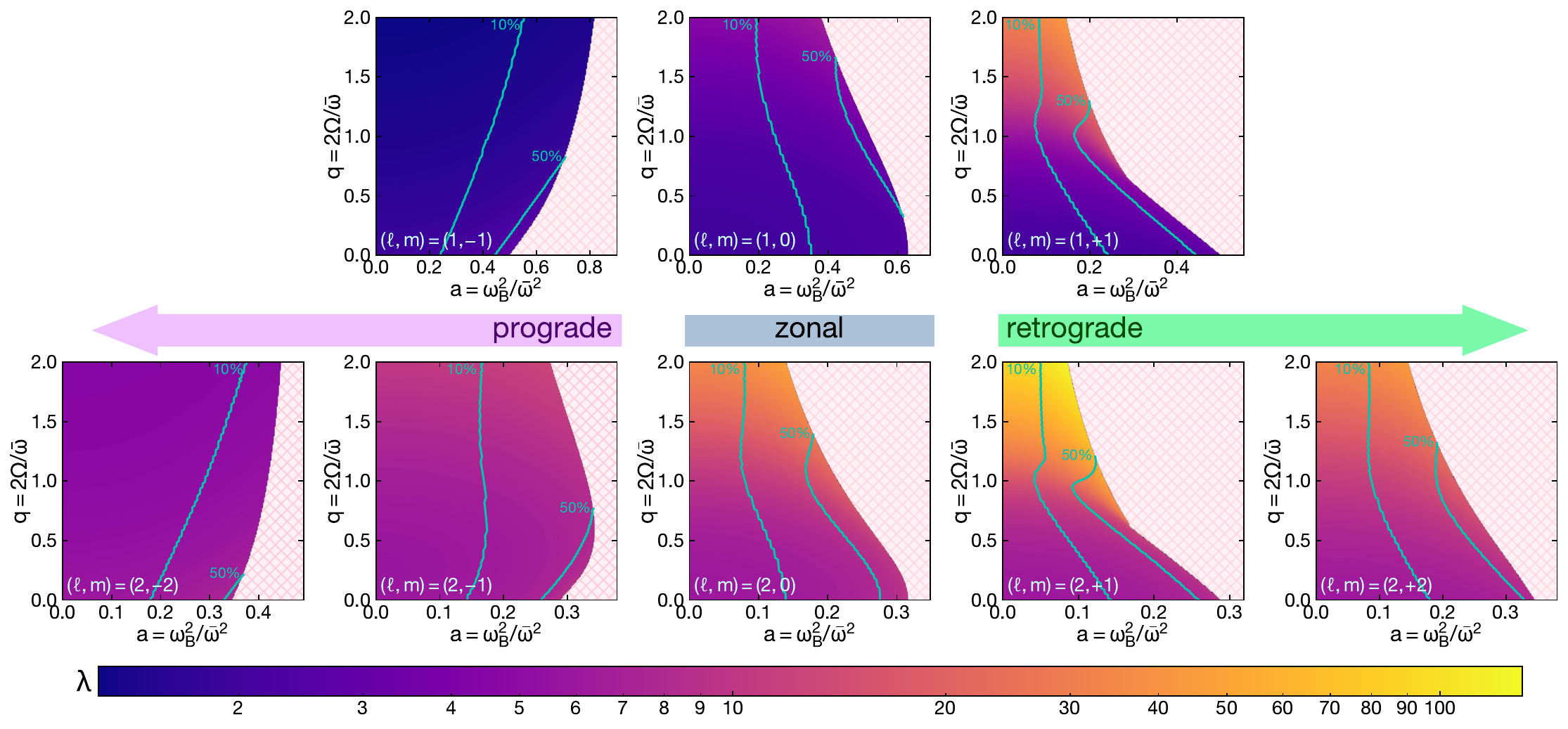}
    \caption{The eigenvalues $\lambda$ of the differential operator $\mathcal{L}^{m,b,q}$ (\autoref{L}) for the dipole (\textit{top}) and quadrupole (\textit{bottom}) modes.
    We plot $\lambda$ against the dimensionless parameters $a=\omega_B^2/\bar{\omega}^2=Nv_{Ar}/r\bar{\omega}^2$ and $q=2\Omega/\bar{\omega}$, which govern the effects of magnetism and rotation, respectively.
    The eigenvalue $\lambda$ enters the asymptotic period spacing as in \autoref{asymptotic}.
    The turquoise lines show contours to the right of which the integrand in \autoref{asymptotic} deviates from a perturbative treatment by 10\% and 50\%.
    The pink hatched zones indicate $b=k_rv_{Ar}/\omega>1$, i.e., modes which occupy these values of $a$ and $q$ at some layer within the star are likely to be suppressed.}.
    \label{fig:parameter_space}
\end{figure*}

\citet{rui2023gravity} calculate numerical solutions to the horizontal problem (\autoref{eig}) in the nonrotating case ($q\!=\!0$) by introducing a small artificial dissipation and using a relaxation scheme.
While this method satisfactorily treats numerical pathologies associated with a singularity at critical latitudes $\mu_\pm\!=\!\pm1/b$ for large fields, it is computationally inefficient.
Relatedly, because the coefficients of \autoref{L} vary quickly across $\mu_\pm$, unreasonably large dissipation coefficients must be assumed to avoid needing prohibitively high resolution near those latitudes.

The more general form of \autoref{eig} that we consider here is still of the Boyd type \citep[e.g.,][]{boyd1981sturm}, but now has solutions, and singular points, indexed by two parameters, $b$ and $q$.
In particular, \autoref{eig} produces four additional singular points, obeying
\begin{equation}
    \mu^2 = \frac{2b^2+q^2 \pm \sqrt{4b^2q^2+q^4}}{2b^4}\mathrm{,}
\end{equation}

\noindent two of which may lie within the solution domain even for fields too weak to resonate with a given oscillation mode (i.e., $b<1$).


We therefore seek an alternative solution strategy that is robust to the presence of such regular singular points.
For $q<1-b^2$, no singularities lie on the domain, and it suffices to perform standard Chebyshev collocation on the real line \citep[e.g.,][]{wang2016computation}.
However, the collocation procedure must be modified somewhat to work for $q>1-b^2$.
We note that since the Sturm--Liouville linear operator in \autoref{L} is analytic, it may be treated as defining an ordinary differential equation on the complex plane.
Solutions to the standard Sturm--Liouville problem on the real line coincide with those of this analytically continued problem, restricted to the real line. Thus, we may construct numerical solutions to the analytically continued problem on a contour on the complex plane, chosen to match the boundary conditions of the real problem on the interval $\mu\!\in\![-1,+1]$. Eigenvalues of the analytically continued problem will not depend on this choice of contour.
Thus, the contour may be chosen to avoid the singular points that we have described above, and therefore to improve the numerical conditioning (e.g. stiffness) of the problem. We refer the reader to, e.g., \citet{boyd_complex_1985} for a more detailed examination of this procedure, and nature of the resulting solutions.

We find the standard collocation procedure to be sufficient for $m\!=\!0$ and $m\!=\!\pm2$ for any values of $q\!\in\![0,2]$ and $b\!\in\![0,1)$.
However, solutions for the $m\!=\!\pm1$ modes under this procedure are numerically badly behaved for $q\!>\!1-b^2$.
In these cases, we perform a complex coordinate transformation from $\mu$ to $\zeta$ given by
\begin{equation}
    \mu = \zeta\left(2-\zeta^2\right) + \zeta\left(1-\zeta^2\right)^2i, \label{eq:contour}
\end{equation}

\noindent and then solve the resulting problem using Chebyshev collocation on the interval $\zeta\!\in\![-1,1]$.
This contour is chosen to share endpoints with the original real interval, while being tangent with the real line from the $|\mu|>1$ (rather than $|\mu|<1$) direction.

The eigenvalue $\lambda$ depends on the relationship between the mode frequency $\bar{\omega}$, rotational frequency $\Omega$ (via $q\!=\!2\Omega/\bar{\omega}$), and magnetogravity frequency $\omega_B$ (via $a\!=\!\omega_B^2/\bar{\omega}^2$).
Therefore, once $\lambda$ is computed for a given pair of $b$ and $q$, we retroactively compute $a=b/\sqrt{\lambda}$), and regard $\lambda$ as being a function of $a$ and $q$.
Because this procedure only produces values of $a$ below some critical $a_{\mathrm{crit}}=a_{\mathrm{crit}}(q)$ (corresponding to the maximum field which permits propagating magnetogravity waves), we excise two families of solutions: the Alfv\'en resonant ones for which $b>1$ \citep[which are expected to experience phase mixing, e.g.,][]{loi2017torsional}, and those which lie on the ``slow'' branch described by \citet{rui2023gravity} (which are expected to approach infinite wavenumber).
Within this work, we consider both such solutions to be ``suppressed'': we do not otherwise make claims about the degree of suppression or the mode frequencies of suppressed modes.




\autoref{fig:parameter_space} shows values of $\lambda$ computed for all dipole ($\ell=1$) and quadrupole ($\ell=2$) modes.
In particular, for the zonal ($m\!=\!0$) and retrograde ($m>0$) modes calculated here, the critical magnetic field needed to cause mode suppression \textit{decreases} with increasing rotation rate.
This is because, for these branches, $\lambda$ increases relatively strongly with $q$ \citep[for $b=0$, $\lambda\propto q^2$;][]{bildsten1996ocean,townsend2003asymptotic,townsend2020improved}.
Therefore, $a_{\mathrm{crit}}=b_{\mathrm{crit}}/\sqrt{\lambda}$ decreases with $q$.
However, since larger rotation rates cause the prograde Kelvin modes (which have $m\!=\!-\ell$) to attain larger horizontal scales ($\lambda$ decreases to a smaller constant value with $q$, when $b=0$), the critical field \textit{increases} with increasing rotation rate.
For the $(\ell,m)=(2,-1)$ case, the dependence of $\lambda$ on $q$ and $b$ is slightly more complicated, hence the non-monotonic behavior of the corresponding critical field with rotation rate.
In any case, a straightforward prediction of this formalism is thus that different branches of modes should undergo suppression at different mode frequencies.
Observational measurements of these critical periods may therefore impose strong constraints on the magnetic and rotational properties of the star.

\section{Numerical solutions of the radial problem} \label{fullradial}

\subsection{Non-asymptotic numerical scheme} \label{nonasymptotic}

In the asymptotic regime, the perturbations vary with radius as $\sim\!e^{i\varphi_g}$, where $\varphi_g$ is given by \autoref{phase} (using the appropriate bounds).
In other words, in this regime, the wavefunctions in the g-mode cavity are expected to be sinusoidal with respect to a modified buoyancy radius $\Pi$ given by
\begin{equation}
    \Pi(r) = \frac{\int^r_{r_1}\frac{N}{r}\,\mathrm{d}r}{\int^{r_2}_{r_1}\frac{N}{r}\,\mathrm{d}r}
\end{equation}

\noindent which we define over the entire main radiative cavity (with respective inner and outer boundaries $r_1$ and $r_2$).
This quantity is normalized such that $\Pi$ ranges from $0$ to $1$.

However, the asymptotic assumption is violated in the proximity of sharp features in $N$ (i.e., buoyancy glitches) when their characteristic widths are $\lesssim k_r^{-1}$.
In such cases, the period spacing is expected to be modified from the asymptotic estimate in \autoref{asymptotic}.
Sharp peaks in $N$ are known to develop at the lower boundaries of the radiative envelopes of evolved MS stars \edit{\citep[in which they cause periodic ``dips'' in the $\delta P_g$ pattern with $P$;][]{miglio2008probing,pedersen2018shape}}, and similar buoyancy glitches have recently been observed asteroseismically in red giants \citep{cunha2015structural,vrard2022evidence}, although their structure is very sensitive to the details of convective boundary mixing \citep[e.g.,][]{michielsen2021probing,lindsay2023near}.
The asymptotic assumption is also strongly violated for g modes with low radial order, which may be observable in subgiants or some pulsators on the MS.

To model some of the non-asymptotic effects, we use a shooting method to solve the stellar pulsation equations under the assumption of adiabaticity, cast in the dimensionless form of \citet{dziembowski1971nonradial}.
This form of the pulsation equations is also employed by commonly used mode-solving codes such as GYRE \citep{townsend2013gyre}.
Rotation and magnetism are implemented only by replacing the angular degree $\ell$ in the equations with an effective degree $\ell_e$, defined in \autoref{effective}.
Thus, we account only for the dynamical effects of rotation and magnetism, and neglect their indirect effects on stellar structure itself.
Additionally, we emphasize that this ``1.5D'' approach still includes both rotation and magnetism asymptotically (similarly to the treatment of rotation in GYRE), and thus relies on the rotation and magnetic field profiles varying slowly in $r$ compared to the wavefunctions themselves.
In other words, while this procedure captures phenomena like wave-trapping due to peaks in $N$, it does \textit{not} accurately model the effects of sharp radial gradients in the magnetic field or rotation profiles, or coupling to, e.g., inertial modes.

In what follows, $M$ and $R$ denote the total mass and radius of the star, and $m$ denotes the mass interior to radius $r$.
We solve the radial problem for
\begin{subequations}
    \begin{gather}
        y_1 = x^{2-\ell_i}\frac{\xi_r}{r} \\
        y_2 = x^{2-\ell_i}\frac{p'}{\rho_0gr}
    \end{gather}
\end{subequations}

\noindent where $x=r/R$, and $\ell_i=\ell_e(r_1)$ is evaluated at the inner boundary.
In buoyancy coordinates, the perturbed time-independent oscillation equations then become
\begin{equation}\label{dziembowski}
    S(\Pi)\frac{\mathrm{d}\vec{y}}{\mathrm{d}\Pi} = \mathbf{A}\vec{y}
\end{equation}

\noindent where
\begin{equation}
    S(\Pi) = \frac{N}{x\int^{r_2}_{r_1}\frac{N}{r}\,\mathrm{d}r}\mathrm{,}
\end{equation}


\noindent $\vec{y}=(y_1,y_2)$, $\sigma=\bar{\omega}\sqrt{R^3/GM}$,
\begin{equation}
    \mathbf{A}
    =
    \begin{pmatrix}
        V/\gamma-1-\ell_e & \lambda/c_1\sigma^2 \\
        c_1\sigma^2 - A^* & 3-U+A^*-\ell_e \\
    \end{pmatrix}\mathrm{,}
\end{equation}


\noindent and
\begin{subequations}
    \begin{gather}
        V = \rho_0rg/p_0 \\
        c_1 = x^3M/m\\
        A^* = N^2r/g \\
        U = 4\pi r^3\rho/m.
    \end{gather}
\end{subequations}

Equations \ref{dziembowski} reflect the $\gamma$-mode localization scheme of \citet{ong2020semianalytic} as well as the Cowling approximation (neglecting perturbations to the gravitational potential).
These approximations are made to restrict our attention to the effect of magnetism and rotation on pure g modes, and to avoid boundary condition-related numerical artifacts \citep[see Section 2.2 of][]{ong2020semianalytic}.
Because the Cowling approximation is well-justified at high radial orders (where the TARM is valid), this approach should capture all of the robust predictions of our formalism.
For the red giant model (Section \ref{models}), the resulting modes should be compared to the output of the stretching procedure typically used to extract $\delta P_g$ from solar-like oscillators \citep{mosser2015period}.

For our numerical shooting, we first integrate \autoref{dziembowski} outwards from the stellar centre as an initial value problem to produce inner basis solutions which are consistent with the boundary conditions imposed there.
In this work, we impose the boundary condition $y_1=0$ ($x_r=0$) on both boundaries.
The solution vector evaluated at any intermediate point (here taken to be $\Pi\!=\!1/2$) should thus be equivalent (up to linear dependence) when obtained by integrating from either boundary (starting from $\vec{y}=(0,1)$).
These two solution vectors \citep[obtained using a Radau integration scheme;][]{wanner1996solving} can then be formed into a $2\times2$ matrix whose determinant $\mathcal{D}(\bar{\omega})$ must vanish at a normal mode $\bar{\omega}=\bar{\omega}^*$.


The adiabatic prescription of \autoref{dziembowski} produces strictly real eigenvalues. To search for modes, we evaluate $\mathcal{D}(\bar{\omega})$ over some frequency grid.
Between frequency grid points where $\mathcal{D}$ changes sign, we use a bisection algorithm to locate the roots of $\mathcal{D}$.
These oscillation modes $\bar{\omega}$ are then converted to their values $\omega$ in the inertial frame via \autoref{converttoinertial} (when there is rotation).



\subsection{Stellar models} \label{models}


\begin{table*}
    \centering
    \begin{tabular}{l l l l l l}
        \hline
        Model & MS-1.5-young & MS-1.5-evolved & RG-1.5 & MS-6.0-young & MS-6.0-evolved \\
        \hline
        $M$ ($M_\odot$) & 1.5 & 1.5 & 1.5 & 6.0 & 6.0 \\
        age & $92\,\mathrm{Myr}$ & $2.49\,\mathrm{Gyr}$ & $2.67\,\mathrm{Gyr}$ & $5\,\mathrm{Myr}$ & $64\,\mathrm{Gyr}$ \\
        $P_{\mathrm{min}}$ (d) & $0.1$ & $0.1$ & $0.026$ ($450\,\mu\mathrm{Hz}$) & $1.0$ & $1.0$ \\
        $P_{\mathrm{max}}$ (d) & $1.0$ & $1.0$ & $0.077$ ($150\,\mu\mathrm{Hz}$) & $3.0$ & $3.0$ \\
        $P_{\mathrm{rot}}$ (d) & $1.5$ & $1.5$ & $30$ & $3.0$ & $3.0$ \\
        $B_c$ (kG) & $475$ & $98$ & $824$ & $172$ & $62$ \\
        evolutionary stage & young MS & evolved MS & red giant & young MS & evolved MS \\
        analogue & young $\gamma$ Dor & evolved $\gamma$ Dor & red giant & young SPB & evolved SPB \\
        results & \autoref{fig:dipole_modes_MS_0} & \autoref{fig:dipole_modes_MS_1} & \autoref{fig:dipole_modes_RG} & \autoref{fig:dipole_modes_MS_2} & \autoref{fig:dipole_modes_MS_3} \\
        \hline
    \end{tabular}
    \caption{Summary of stellar models for which we calculate oscillation modes using the non-asymptotic scheme described in Section \ref{nonasymptotic}.}
    \label{tab:models}
\end{table*}

We find the oscillation modes of stellar models produced using version r22.11.1 of the Modules for Stellar Experiments (MESA) code \citep{paxton2010modules,paxton2013modules,paxton2015modules,paxton2018modules,paxton2019modules}.
We incorporate realistic convective overshoot using exponential overmixing with scale height $f_{\mathrm{ov}}H_p=0.015H_p$ (where $H_p$ is the local pressure scale height), with the overshooting region starting a distance $0.005H_p$ inside the convective zone.

The stellar profiles as well as the rotation periods and magnetic fields we assume for them are summarized in \autoref{tab:models}.
In particular, we choose three snapshots from a $1.5M_\odot$ model to assess the behavior of the period spacing on the early-MS (MS-1.5-young), late-MS (MS-1.5-evolved), and lower RG (RG-1.5), and two snapshots from a $6.0M_\odot$ star on the early-MS (MS-6.0-young) and late-MS (MS-6.0-evolved).
These models are chosen to be representative of $\gamma$ Dor (MS-1.5-young, MS-1.5-evolved), slowly pulsating B-type (SPB; MS-6.0-young, MS-6.0-evolved), and red giant solar-like (RG-1.5) oscillators.
We solve for the dipole ($\ell=1$) oscillation modes over a realistic range of frequencies.
For RG-1.5, we compute these frequencies with both rotation and magnetism, as well as in the absence of either, in order to benchmark the prediction of perturbation theory (\autoref{redgiant}).
For the main-sequence models, the mode frequencies are computed three times, including the effects of magnetism and rotation both separately and simultaneously (\autoref{msresults}).
The magnetic field is chosen to be strong enough to exhibit the effects of strong magnetic modification and suppression of some branch of oscillation modes.
The mode period/frequency ranges shown in \autoref{tab:models} are given in the inertial frame.
When relevant, we solve only for co-rotating frequencies $\bar{\omega}\gtrsim0$ to avoid the pile-up of g modes close to $\bar{\omega}=0$.

Our models do not take into account distortions of the stellar structure due to centrifugal forces and magnetic pressure.
While these effects are unlikely to matter in most observed $\gamma$ Dor and SPB stars \citep{henneco2021effect}, they are likely to be important in rapidly rotating p-mode pulsators \citep[such as $\delta$ Sct stars, e.g.,][]{lignieres2006acoustic}.

\section{Results and discussion} \label{results}

\subsection{Strong fields in red giant cores} \label{redgiant}

Strong magnetic fields in red giant cores have two main asteroseismic manifestations.
First, they may produce frequency shifts on the nonradial modes which tend to shift modes of all $m$ in the same direction (as opposed to rotation, which creates a frequency multiplet).
Measurements of such frequency shifts have recently been used to make inferences about the field strength and, in one case, even geometry \citep{li202230,li2023internal}.
Second, if the magnetic field is extraordinarily strong, the magnetic field is expected to suppress the amplitudes of dipole modes whose frequencies lie below some $\omega_{\mathrm{crit}}\!\sim\!\omega_B$ \citep{fuller2015asteroseismology,lecoanet2017conversion,rui2023gravity}.

\begin{figure}
    \centering
    \includegraphics[width=0.46\textwidth]{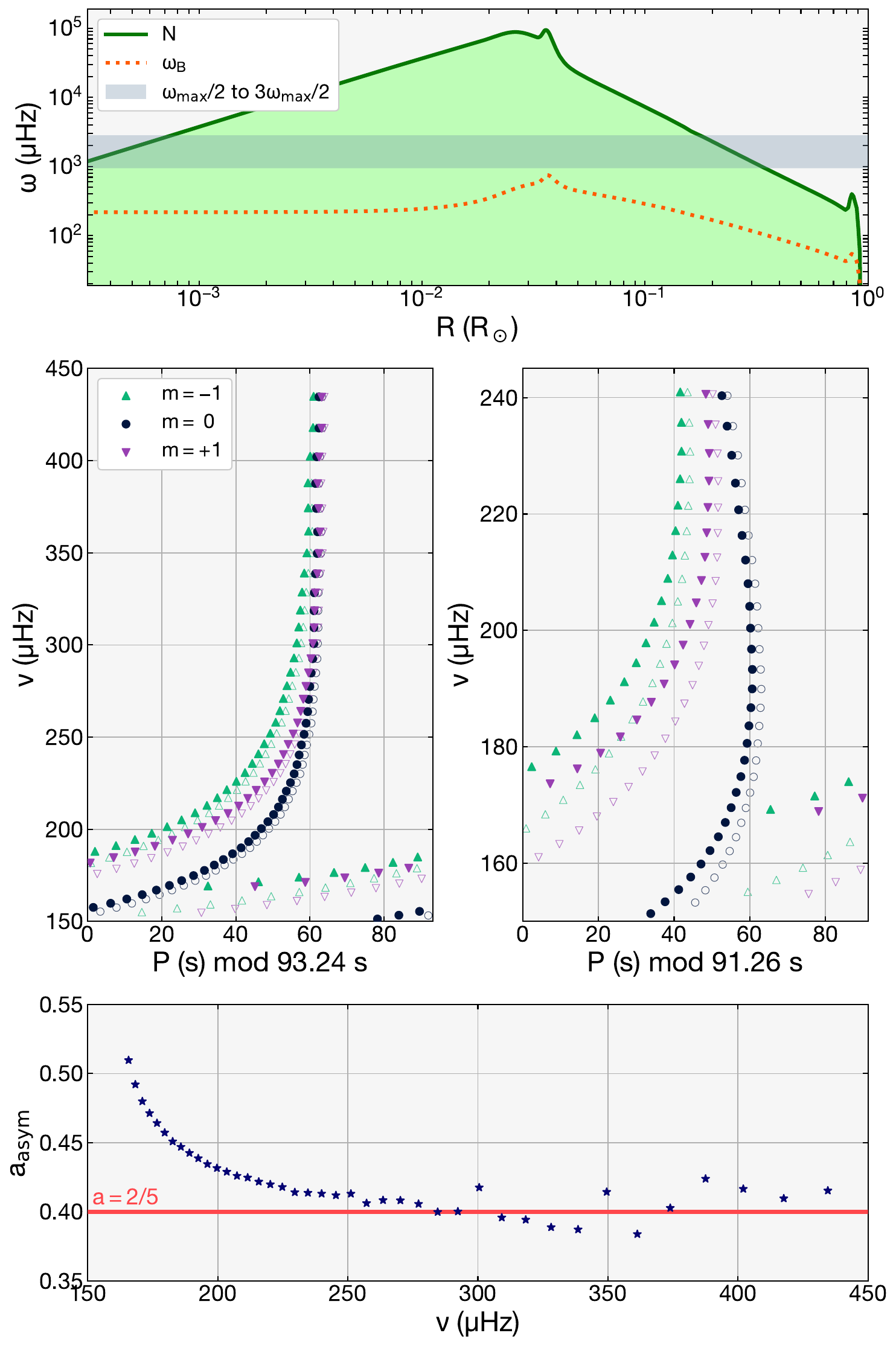}
    \caption{\textit{Top:} The Brunt--V\"ais\"al\"a ($N$) and magnetogravity ($\omega_B$) frequencies for the red giant model (RG-1.5), plotted in relation to the range over which we solve for mode frequencies.
    The rotational frequency $\Omega\simeq2.4\,\mu\mathrm{Hz}$ ($P_{\mathrm{rot}}=30\,\mathrm{d}$) is below the bottom bound of this plot.
    \textit{Center:} Period echelle diagram for the red giant's core g modes.
    The right panel zooms into the low frequency modes of the left panel, and folds on a different period for clarity.
    Solid symbols denote mode frequencies calculated using the TARM, whereas hollow symbols denote the lowest-order prediction of perturbation theory.
    \textit{Bottom:} The dipole asymmetry parameter (\autoref{asymmetry}) plotted against unperturbed mode frequency.}
    \label{fig:dipole_modes_RG}
\end{figure}

Our red giant model (RG-1.5; described in Section \ref{models}) is chosen to mimic a star on the lower red giant branch (for which mixed modes are easiest to observe) with a typical rotation rate ($P_{\mathrm{rot}}\!=\!30\,\mathrm{d}$).
For a frequency of maximum power $\nu_{\mathrm{max}}\!\approx\!300\,\mu\mathrm{Hz}$, we calculate all dipole modes within the frequency range $\nu_{\mathrm{max}}/2$ and $3\nu_{\mathrm{max}}/2$ in the simultaneous presence of magnetism and rotation, using the scheme described in Section \ref{nonasymptotic}.
The width of the adopted frequency range is comparable to the full width at half maximum value $\delta\nu_{\mathrm{env}}\!\approx\!100\,\mu\mathrm{Hz}\approx\nu_{\mathrm{max}}/3$ calculated using the scaling relation of \citet{mosser2012characterization}.
The large central magnetic field $B_c\!\approx\!820\,\mathrm{kG}$ is chosen such that the $m\!=\!\pm 1$ sectoral modes are suppressed at the lower frequency range, to show the effect of a strong field.
Note that $B_c$ refers to the maximum value of the radial component of the field at the center of the star, rather than some horizontally averaged version of this quantity.
Therefore, this value of $B_c$ corresponds to a horizontally averaged field $\overline{B_r^2}^{1/2}\!\!\approx\!B_c/\sqrt{3}\!\approx\!470\,\mathrm{kG}$ when normalized in the same way as the values reported by \citet{li202230} ($30$--$100\,\mathrm{kG}$), \citet{deheuvels2023strong} ($40$--$610\,\mathrm{kG}$), and \citet{li2023internal} ($20$--$150\,\mathrm{kG}$).
The middle panels of \autoref{fig:dipole_modes_RG} show mock period echelle diagrams corresponding to these calculations.

We additionally calculate the mode frequencies for the same stellar model in the absence of either rotation and magnetism, in order to test the perturbative formalism.
At high frequencies (where both rotation and magnetism are perturbative), the mode frequencies are closely consistent with the perturbative frequency shifts derived by \citet{li202230} (the \edit{unfilled symbols} in \edit{the middle panels of} \autoref{fig:dipole_modes_RG}).
However, at low frequencies close to suppression ($\nu\lesssim220\,\mu\mathrm{Hz}$), the TARM and perturbative results deviate substantially, with the TARM results tending to predict much larger frequency shifts than the perturbative formulae.
This effect becomes increasingly dramatic until, at $\nu\approx170\,\mu\mathrm{Hz}$, the sectoral modes are totally suppressed (although the zonal $m\!=\!0$ mode remains propagating, and is suppressed at a frequency below the chosen observed frequency range).
Disagreement between the perturbative and TARM frequency shifts is fully expected: at or near suppression, the effects of magnetism are, by definition, highly non-perturbative.

To formally demonstrate consistency with the perturbative formulae at high mode frequencies, we can expand the operator in \autoref{L} in $b$ and $q$ and perform a perturbation analysis.
Corrections to the subsequent analysis enter at $\mathcal{O}(q^3,qa^2,a^4)\sim\mathcal{O}(\Omega^3,\Omega\omega_B^4,\omega_B^8)$.
We obtain the following eigenvalue equation:
\begin{equation} \label{perted}
    0 = \lambda p'(\mu) + \mathcal{L}^m_0[p'(\mu)] + \mathcal{L}^{m,b,q}_{\mathrm{pert}}[p'(\mu)]
\end{equation}

\noindent where
\begin{equation}
    \begin{split}
        \mathcal{L}^{m,b,q}_{\mathrm{pert}}[p'(\mu)] &= -mqp'(\mu) \\
        &+ (b^2+q^2)\left[\frac{\mathrm{d}}{\mathrm{d}\mu}\left(\mu^2(1-\mu^2)\frac{\mathrm{d}p'}{\mathrm{d}\mu}\right) - \frac{m^2\mu^2}{1-\mu^2}p'(\mu)\right]\mathrm{.}
    \end{split}
\end{equation}

To find the effect of $\mathcal{L}^{m,b,q}_{\mathrm{pert}}$ on the eigenvalues, we perform first-order perturbation theory.
If the dipole eigenvalues are given by
\begin{equation}
    \lambda = \ell(\ell+1) + \tilde{\lambda}^{m,b,q} = 2 + \tilde{\lambda}^{m,b,q}\mathrm{,}
\end{equation}

\noindent where
\begin{equation} \label{firstorderpert}
    0 = \tilde{\lambda}^{m,b,q} + \int^{+1}_{-1}p'^0_m(\mu)^*\mathcal{L}^{m,b,q}_{\mathrm{pert}}[p'^0_m(\mu)]\,\mathrm{d}\mu
\end{equation}

\noindent and $p'^0_m(\mu)$ are the unperturbed eigenfunctions (of $\mathcal{L}^m_0$).
We emphasize that this is a perturbative expansion on the space of latitudinal functions all of the same $m$ (for the generalized Legendre operators), \edit{and \textit{not} on the full space of spherical harmonics \citep[as done by][]{li202230}.}
Degenerate perturbation theory is thus not necessary here\edit{, since the eigenvalues of the generalized Legendre operator for a given $m$ do not repeat}.
Furthermore, while in principle corrections may enter in an expression at second-order in perturbation theory, the only relevant term $\propto\!-mq$ in $\mathcal{L}^{m,b,q}_{\mathrm{pert}}$ shifts all of the eigenvalues of a given $m$ equally, and thus does not induce a second-order perturbation in $\lambda$.

The unperturbed pressure perturbations are the associated Legendre polynomials,
\begin{equation} \label{nopertp}
    p'^0_m(\mu)
    =
    \begin{cases}
        \sqrt{\frac{3}{2}}\mu & m\!=\!0 \\
        \sqrt{\frac{3}{4}}\sqrt{1-\mu^2} & m\!=\!\pm1
    \end{cases}
\end{equation}

\noindent where we have normalized the functions to square-integrate to unity and ignored the overall (Condon--Shortley) phase.
The integral in \autoref{firstorderpert} can therefore be evaluated to give
\begin{equation}
    \lambda
    =
    \begin{cases}
        2 + \frac{2}{5}(b^2+q^2) & m\!=\!0 \\
        2 \pm q + \frac{4}{5}(b^2+q^2) & m\!=\!\pm 1
    \end{cases}
\end{equation}

\noindent so that
\begin{equation}
    \sqrt{\lambda}
    =
    \begin{cases}
        \sqrt{2} + \frac{1}{5\sqrt{2}}(b^2+q^2) & m\!=\!0 \\
        \sqrt{2} \pm \frac{1}{2\sqrt{2}}q + \frac{1}{80\sqrt{2}}(32b^2+27q^2) & m\!=\!\pm 1
    \end{cases}
    \mathrm{.}
\end{equation}

To transform the independent variable $b$ to $a$ (which can directly be specified given a field and stellar profile), we note that
\begin{equation}
    b = a\sqrt{\lambda} \approx a\sqrt{2}
\end{equation}

\noindent up to the relevant order.
Then
\begin{equation}
    \sqrt{\lambda}
    =
    \begin{cases}
        \sqrt{2} + \frac{1}{5\sqrt{2}}(2a^2+q^2) & m\!=\!0 \\
        \sqrt{2} \pm \frac{1}{2\sqrt{2}}q + \frac{1}{80\sqrt{2}}(64a^2+27q^2) & m\!=\!\pm 1
    \end{cases}
    \mathrm{.}
\end{equation}

The mode frequencies in the co-rotating frame are given by
\begin{equation} \label{ugly}
    \bar{\omega} = \frac{1}{\varphi_g}\int^{r_2}_{r_1}\sqrt{\lambda}\,\frac{N}{r}\,\mathrm{d}r
\end{equation}

\noindent in the asymptotic regime, where \edit{$\varphi_g=\pi(n_g+\varepsilon_g)$ \citep{tassoul} is the total radial phase (note that $\lambda$ depends implicitly on $\bar{\omega}$ in a complicated way).}
We again proceed in ignoring the frequency dependence of the bounds of the integral in \autoref{ugly} (which should formally only enclose the part of the main radiative cavity where $\omega<N$).



We define the ``buoyant average'':
\begin{equation}
    \langle\ldots\rangle_g = \frac{\int^{r_2}_{r_1}(\ldots)\frac{N}{r}\,\mathrm{d}r}{\int^{r_2}_{r_1}\frac{N}{r}\,\mathrm{d}r}\mathrm{.}
\end{equation}

Assuming that $\delta\bar{\omega}\!\ll\!\bar{\omega}_0$ (sufficient for the desired order of the expansion), we may expand \autoref{ugly} as
\begin{equation} \label{complicated1}
    \delta\bar{\omega} = \left(\frac{\langle\omega_B^4\rangle_g}{5\bar{\omega}_0^3} + \frac{2\langle\Omega^2\rangle_g}{5\bar{\omega}_0}\right) - \sqrt{2}\left(\frac{2\langle\omega_B^4\rangle_g}{5\bar{\omega}_0^3}+\frac{2\langle\Omega^2\rangle_g}{5\bar{\omega}_0}\right)\frac{\delta\bar{\omega}}{\bar{\omega}_0}
\end{equation}

\noindent for $m\!=\!0$, and
\begin{equation} \label{complicated2}
    \begin{split}
        \delta\bar{\omega} = &\left(\pm \frac{\langle\Omega\rangle_g}{2}+\frac{2\langle\omega_B^4\rangle_g}{5\bar{\omega}_0^3}+\frac{27\langle\Omega^2\rangle_g}{40\bar{\omega}_0}\right)\\
        &- \sqrt{2}\left(\pm\frac{\langle\Omega\rangle_g}{4} - \frac{8\langle\omega_B^4\rangle_g}{5\bar{\omega}_0^3} - \frac{27\langle\Omega^2\rangle_g}{40\bar{\omega}_0}\right)\frac{\delta\bar{\omega}}{\bar{\omega}_0}
    \end{split}
\end{equation}

\noindent for $m\!=\!\pm1$.
Equations \ref{complicated1} and \ref{complicated2} can be solved to yield the following frequency shifts:

\begin{subequations} \label{freqshifts}
    \begin{gather}
        \delta\bar{\omega}_{m=0} = \frac{1}{5}\frac{\langle\omega_B^4\rangle_g}{\bar{\omega}_0^3} + \frac{2}{5}\frac{\langle\Omega^2\rangle_g}{\bar{\omega}_0} \\
        \delta\bar{\omega}_{m=\pm1} = \pm\frac{\langle\Omega\rangle_g}{2} + \frac{2}{5}\frac{\langle\omega_B^4\rangle_g}{\bar{\omega}_0^3} + \frac{27\langle\Omega^2\rangle_g-10\langle\Omega\rangle_g^2}{40\bar{\omega}_0}\mathrm{.}
    \end{gather}
\end{subequations}

We keep one higher order of the rotation rate than do \citet{li202230}.
We distinguish between $\langle\Omega^2\rangle_g$ and $\langle\Omega\rangle_g^2$ in the above to allow for the possibility of weak differential rotation \citep[e.g.,][]{beck2012fast}, which may distinguish between the two.
However, in the case of uniform rotation (assumed throughout this work), $\langle\Omega^2\rangle_g=\langle\Omega\rangle_g^2=\Omega^2$.
In the inertial frame, these frequency shifts become
\begin{subequations} \label{freqshiftsinertial}
    \begin{gather}
        \delta\omega_{m=0} = \frac{1}{5}\frac{\langle\omega_B^4\rangle_g}{\omega_0^3} + \frac{2}{5}\frac{\langle\Omega^2\rangle_g}{\omega_0} \\
        \delta\bar{\omega}_{m=\pm1} = \mp\frac{\langle\Omega\rangle_g}{2} + \frac{2}{5}\frac{\langle\omega_B^4\rangle_g}{\omega_0^3} + \frac{27\langle\Omega^2\rangle_g-10\langle\Omega\rangle_g^2}{40\omega_0}
    \end{gather}
\end{subequations}

\noindent where $\bar{\omega}_0=\omega_0$ for the unperturbed modes.
We have full consistency with the perturbation formulae of \citet{li202230} (their Equations 61 and 62, with $\zeta=1$).
Note that the star-averaged quantity which they define to be $\omega_B$ ($\equiv\omega_{B,\mathrm{L22}}$) is equal to $\omega_{B,\mathrm{L22}}=\langle\omega_B^4\rangle_g/3\bar{\omega}_0^3$.

We caution that both the direct role of the centrifugal force as a restorative force and its indirect impact on the stellar structure \citep[e.g.,][]{ballot2010gravity} also enter at $\propto\Omega^2$.
Inclusion of these effects is likely necessary to accurately capture the second-order effects of rotation.

Our non-perturbative mode calculations imply a few straightforward predictions.
First, as mentioned previously, the magnetic frequency shifts become substantially stronger than implied by a perturbative estimate.
While the relative change in the period spacing is still small ($\delta P_g$ decreases by $\approx10\%$ before suppression), the frequency shifts still substantially modify the period echelle diagram.
Conversely, if the period spacing pattern of a strongly magnetic red giant is fit using the perturbative formulae, the inferred magnetic field is likely to be a significant overestimate.
For example, \citet{deheuvels2023strong} claim the detection of a red giant (KIC 6975038) whose magnetic field ($\approx286\,\mathrm{kG}$) significantly exceeds the critical field $B_{\mathrm{crit}}$ by a factor $\sim1.7$.
Under our formalism, a field near or exceeding $B_{\mathrm{crit}}$ should efficiently damp magnetogravity waves, either through phase mixing or refraction to infinite wavenumbers.
Indeed, \citet{deheuvels2023strong} observe nearly total dipole suppression in the same star for only low-frequency modes $\lesssim\!\nu_{\mathrm{max}}$, consistent with $\omega_{\mathrm{crit}}$ lying on their observed frequency range.
Their results could potentially be brought into accord with ours if non-perturbative effects have caused an observational overestimate of the field by a factor of a few.

\begin{figure}
    \centering
    \includegraphics[width=0.45\textwidth]{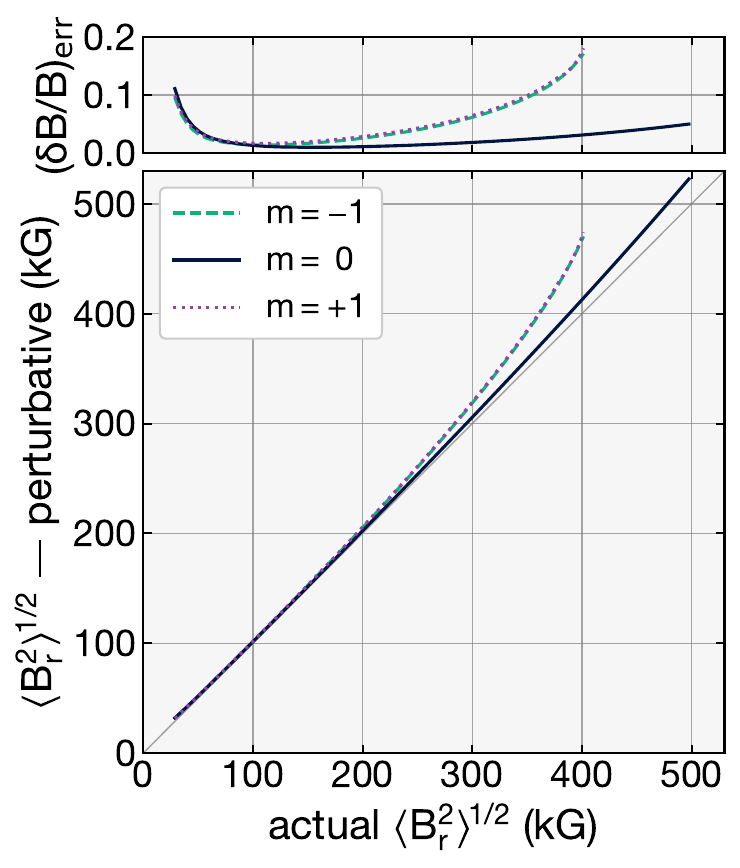}
    \caption{For the red giant model (RG-1.5): \textit{Top:} The relative error on the inferred magnetic field $\langle B_r^2\rangle^{1/2}$ associated with perturbation theory.
    $\langle B_r^2\rangle^{1/2}$ refers to an angle- and radial kernel-averaged field, following the notation of \citet{li202230} (see \autoref{averaged}).
    \textit{Bottom:} The internal magnetic field $\langle B_r^2\rangle^{1/2}$ implied by perturbation theory using the frequency shift for some angular degree $m$, plotted against the ``real'' value (given by our non-perturbative TARM formalism).
    The frequency shift is evaluated \edit{using $\varphi_g\simeq\pi n_g$ for a physically realistic radial order $n_g=70$} ($\nu_{\mathrm{max}}\approx150\,\mu\mathrm{Hz}$), roughly the bottom of the frequency range shown in the period echelle diagram in \autoref{fig:dipole_modes_RG}.}
    \label{fig:show_overestimated_field}
\end{figure}

To characterize the severity of such systematic overestimates, we compute the dipole frequency shifts in the red giant model for a range of internal magnetic fields (by numerically solving \autoref{ugly}).
For each order $m$, we then calculate the internal magnetic field which would be needed to produce the same frequency shift in perturbation theory.
\autoref{fig:show_overestimated_field} shows that the magnetic field $\langle B_r^2\rangle^{1/2}$ implied by perturbation theory can exceed the ``true'' value for fields which are almost strong enough to cause suppression.
Specifically, we use $\langle B_r^2\rangle^{1/2}$ to denote the field averaged over all angles and over the radial kernel \citep[following][]{li202230}:
\begin{equation} \label{averaged}
    \langle B_r^2\rangle = \frac{1}{3}\int^{r_2}_{r_1}K(r)B_r^2\,\mathrm{d}r.
\end{equation}

\noindent where $K(r)$ is given by \autoref{ker} in the asymptotic limit.

While the errors accrued by the perturbative formulae in \autoref{fig:show_overestimated_field} are relatively small and do not rise to a factor $\sim\!1.7$, the degree to which perturbation theory overestimates the field likely depends on the field geometry adopted and the exact structure of the star (via, e.g., how far up the red giant branch the star is).
Moreover, it likely depends on the exact procedure used to extract the field.
For example, \autoref{fig:show_overestimated_field} shows magnetic field values inferred using only one azimuthal order at one frequency, but an inference using the whole oscillation spectrum may yield a different answer.
In the future, the manner in which perturbation theory misestimates the field should be characterized in more detail as a function of these factors.
Large relative errors in the inferred magnetic field may also appear at low fields end if second-order rotational effects are mistaken for magnetic shifts (top panel of \autoref{fig:show_overestimated_field}).

Second, \citet{li202230} and \citet{li2023internal} measure the dipole asymmetry parameter, defined by
\begin{equation} \label{asymmetry}
    \begin{split}
        a_{\mathrm{asym}} &= \frac{\delta\omega_{m=+1}+\delta\omega_{m=-1}-2\delta\omega_{m=0}}{\delta\omega_{m=+1}+\delta\omega_{m=-1}+\delta\omega_{m=0}}\mathrm{.}
    \end{split}
\end{equation}

This should not be confused with the parameter $a\!=\!\omega_B^2/\omega^2$ defined in this work and by \citet{rui2023gravity}.
In the perturbative regime, they show that
\begin{equation} \label{asymphys}
    a_{\mathrm{asym}} = \frac{\int^{r_2}_{r_1}\mathrm{d}r\,K(r)\iint\sin\theta\,\mathrm{d}\theta\,\mathrm{d}\phi\,B_{0r}^2P_2(\cos\theta)}{\int^{r_2}_{r_1}\mathrm{d}r\,K(r)\iint\sin\theta\,\mathrm{d}\theta\,\mathrm{d}\phi\,B_{0r}^2}\mathrm{,}
\end{equation}

\noindent where $P_2(\mu)=(3\mu^2-1)/2$ is the second-order Legendre polynomial and $K(r)$ is a radial kernel function given by
\begin{equation} \label{ker}
    K(r) = \frac{\rho^{-1}(N/r)^3}{\int^{r_2}_{r_1}\rho^{-1}(N/r)^3\,\mathrm{d}r}\mathrm{.}
\end{equation}

In particular, when the horizontal dependence of $B_{0r}$ is given by $\psi(\theta,\phi)$ (i.e., the horizontal geometry is radius-independent), the radial integral in \autoref{asymphys} can be eliminated, yielding
\begin{equation} \label{asymphys2}
    a_{\mathrm{asym}} = \frac{\iint\sin\theta\,\mathrm{d}\theta\,\mathrm{d}\phi\,\psi(\theta,\phi)^2P_2(\cos\theta)}{\iint\sin\theta\,\mathrm{d}\theta\,\mathrm{d}\phi\,\psi(\theta,\phi)^2}\mathrm{.}
\end{equation}

In the special case of a dipole magnetic field whose axis is aligned with the rotational axis ($\psi(\theta,\phi)=\cos\theta$), it can be seen that $a_{\mathrm{asym}}=2/5=0.4$ in this expression.

In the bottom panel of \autoref{fig:dipole_modes_RG}, we see that this expectation holds at high frequencies, but increases slightly to $\approx\!0.5$ at lower frequencies (near $\omega_{\mathrm{crit}}$).
While likely difficult to measure, a value of $a_{\mathrm{asym}}$ that varies towards lower frequencies (coinciding with the inference of a large magnetic field from the frequency shifts) may be an independent signature of a near-critical field.
This non-perturbative asymmetry effect is related to the different magnetic fields implied by perturbation theory's predictions for the frequency shifts of different azimuthal orders (\autoref{fig:show_overestimated_field}).


In stars with especially weak magnetic fields, it is in principle possible for the dipole asymmetry to be dominated by rotation, \textit{even if} it is slow enough for perturbation theory to be applicable.
From \autoref{freqshiftsinertial} and Equations \ref{asymmetry}, we have
\begin{equation} \label{asym}
    a_{\mathrm{asym}} = \frac{8\langle\omega_B^4\rangle_g/\bar{\omega}_0^2+11\langle\Omega^2\rangle_g-10\langle\Omega\rangle_g^2}{20\langle\omega_B^4\rangle_g/\bar{\omega}_0^2+35\langle\Omega^2\rangle_g-10\langle\Omega\rangle_g^2}
\end{equation}

\noindent such that, for a uniform rotation rate $\Omega$,
$\langle\Omega^2\rangle_g=\langle\Omega\rangle_g^2=\Omega^2$), \autoref{asym} possesses the limiting behavior
\begin{equation}
    a_{\mathrm{asym}}
    =
    \begin{cases}
        \frac{2}{5} - \frac{9}{20}\frac{\bar{\omega}_0^2\Omega^2}{\langle\omega_B^4\rangle_g} & \langle\omega_B^4\rangle_g\gg\bar{\omega}_0^2\Omega^2 \\
        \frac{1}{25} + \edit{\frac{36}{125}\frac{\langle\omega_B^4\rangle_g}{\bar{\omega}_0^2\Omega^2}} & \langle\omega_B^4\rangle_g\ll\bar{\omega}_0^2\Omega^2
    \end{cases}
    \mathrm{.}
\end{equation}

When the magnetic asymmetry dominates ($\langle\omega_B^4\rangle_g/\bar{\omega}_0^2\gg\langle\Omega\rangle_g^2\simeq\Omega^2$), $a_{\mathrm{asym}}\!\approx\!2/5\!=\!0.40$.
However, when the Coriolis-induced rotational asymmetry dominates ($\Omega^2\gg\langle\omega_B^4\rangle_g/\bar{\omega}_0^2$), we instead have $a_{\mathrm{asym}}\approx1/25=0.04$.
We stress that this is a fully perturbative effect: it only deviates from the result of \citet{li202230} because it includes a single higher-order effect of rotation.
The upshot is that, even when both rotation and magnetism are individually small, $a_{\mathrm{asym}}\neq2/5$ for aligned rotational/magnetic axes if the effect of rotation is \textit{relatively} at least comparable to that of magnetism.
We again caution that the centrifugal force (which is also relevant at this order in $\Omega$) has been neglected---this likely implies that the rotation-dominated asymmetry does not exactly approach $1/25$ but some other value.
Inclusion of such effects \citep[as done by, e.g.,][]{mathis2019traditional,dhouib2021traditional,dhouib2021traditional2} is needed to properly predict the true rotation-dominated asymmetry value.
Nevertheless, we expect the qualitative ability for rotation to dominate over magnetism in determining the dipole asymmetry to be robust.

\citet{li202230} and \citet{li2023internal} neglect the rotational asymmetry effect on the basis that the core rotation rates in the stars in their sample are typical (i.e., low): we hereafter check this explicitly.
As a crude estimate, the magnetic asymmetry dominates the rotational asymmetry in a red giant core when $\langle\omega_B^4\rangle_g/\omega_{\mathrm{max}}^2\Omega^2\gg1$.
In the three stars investigated by \citet{li202230}, $\langle\omega_B^4\rangle_g/\omega_{\mathrm{max}}^2\Omega^2\gtrsim10^2$ and their asymmetries are thus indeed very magnetically dominated.
Most of the stars reported by \citet{li2023internal} have values of $\langle\omega_B^4\rangle_g/\omega_{\mathrm{max}}^2$ in the tens or hundreds.
However, this parameter reaches a minimum for KIC 8540034, for which $\langle\omega_B^4\rangle_g/\omega_{\mathrm{max}}^2\Omega^2\approx9$.
In this star, rotation may affect the asymmetry parameter for low-frequency modes (note the frequency dependence of $\langle\omega_B^4\rangle_g/\omega^2\Omega^2$)
In general, magnetic domination of the dipole asymmetry may not be the case for giants with either fast core rotation rates \textit{or} weak fields, and we caution against using $a_{\mathrm{asym}}$ alone to make an inference of the field geometry without checking this criterion explicitly.

\subsection{Strong fields threading the envelopes of main-sequence pulsators} \label{msresults}

\begin{figure}
    \centering
    \includegraphics[width=0.46\textwidth]{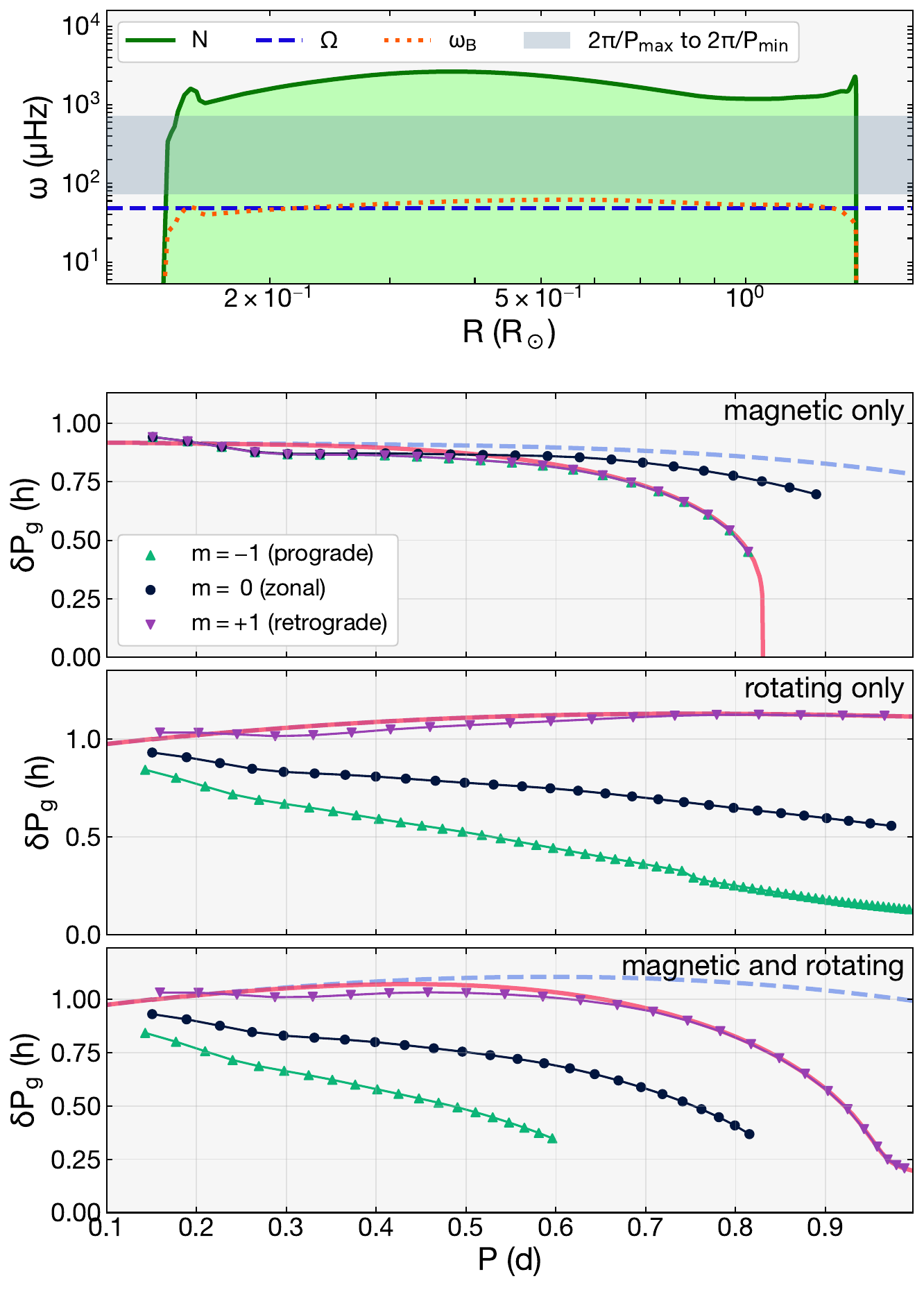}
    \caption{Characteristic frequency profiles and mode frequencies for a young $\gamma$ Dor analogue (MS-1.5-young).
    \textit{Top:} The Brunt--V\"ais\"al\"a ($N$), rotational ($\Omega$), and magnetogravity ($\omega_B$) frequencies, plotted in relation to the range over which we solve for mode frequencies.
    \textit{Bottom:} The period spacing $\delta P_g$ versus period $P$ in the inertial frame for the dipole modes, in the magnetic, rotating, and magnetic and rotating cases.
    Predictions for the asymptotic period spacing for the $m\!=\!1$ branch (using \autoref{asymptotic}) are shown in \textit{solid red}.
    We also show predictions for the asymptotic period spacing handling rotation non-perturbatively but magnetism only perturbatively (using \autoref{pertmagonly}; \textit{dashed blue} curves).}
    \label{fig:dipole_modes_MS_0}
\end{figure}

\begin{figure}
    \centering
    \includegraphics[width=0.46\textwidth]{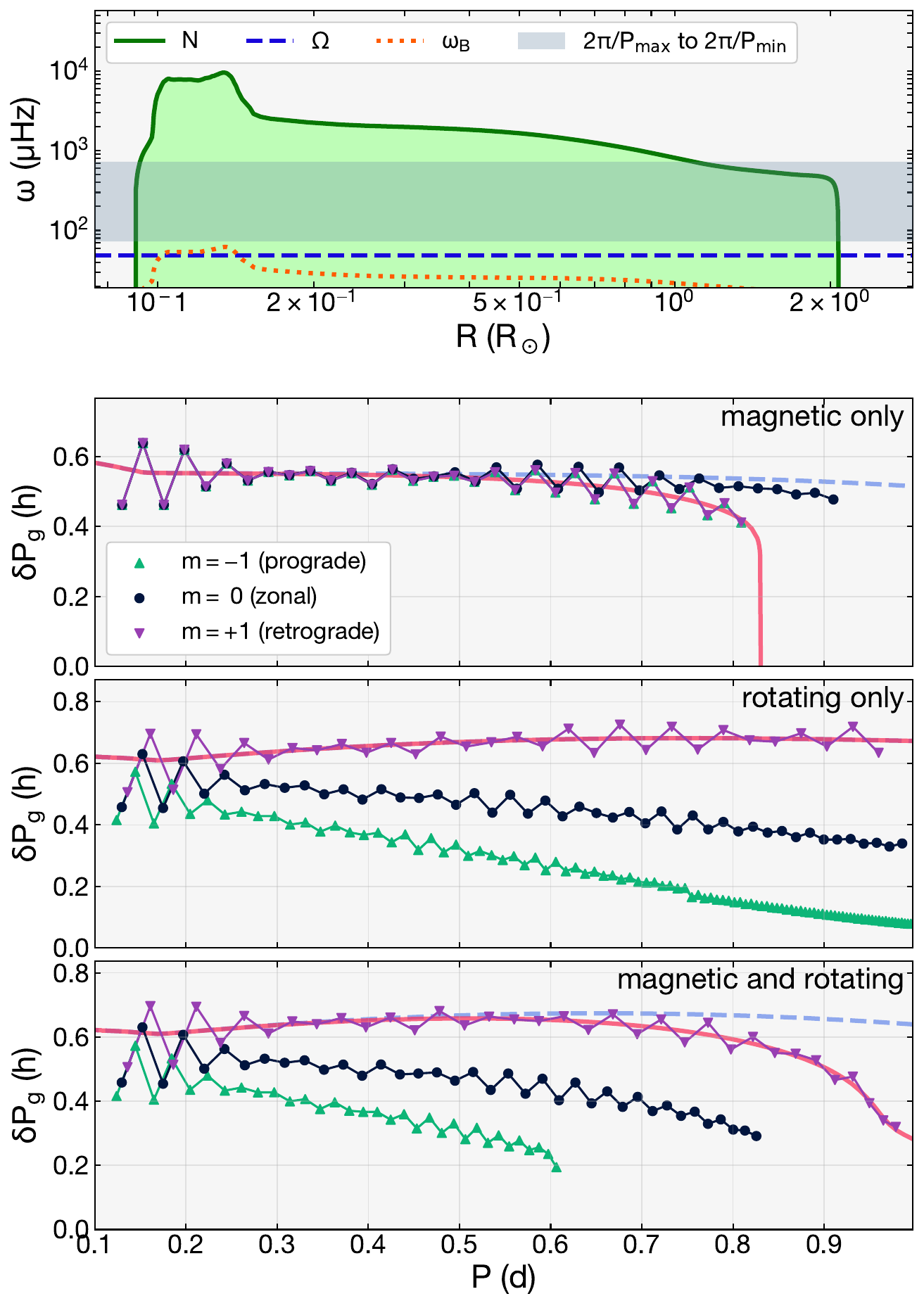}
    \caption{Same as \autoref{fig:dipole_modes_MS_0}, but for an evolved $\gamma$ Dor analogue (MS-1.5-evolved).}
    \label{fig:dipole_modes_MS_1}
\end{figure}

\begin{figure}
    \centering
    \includegraphics[width=0.46\textwidth]{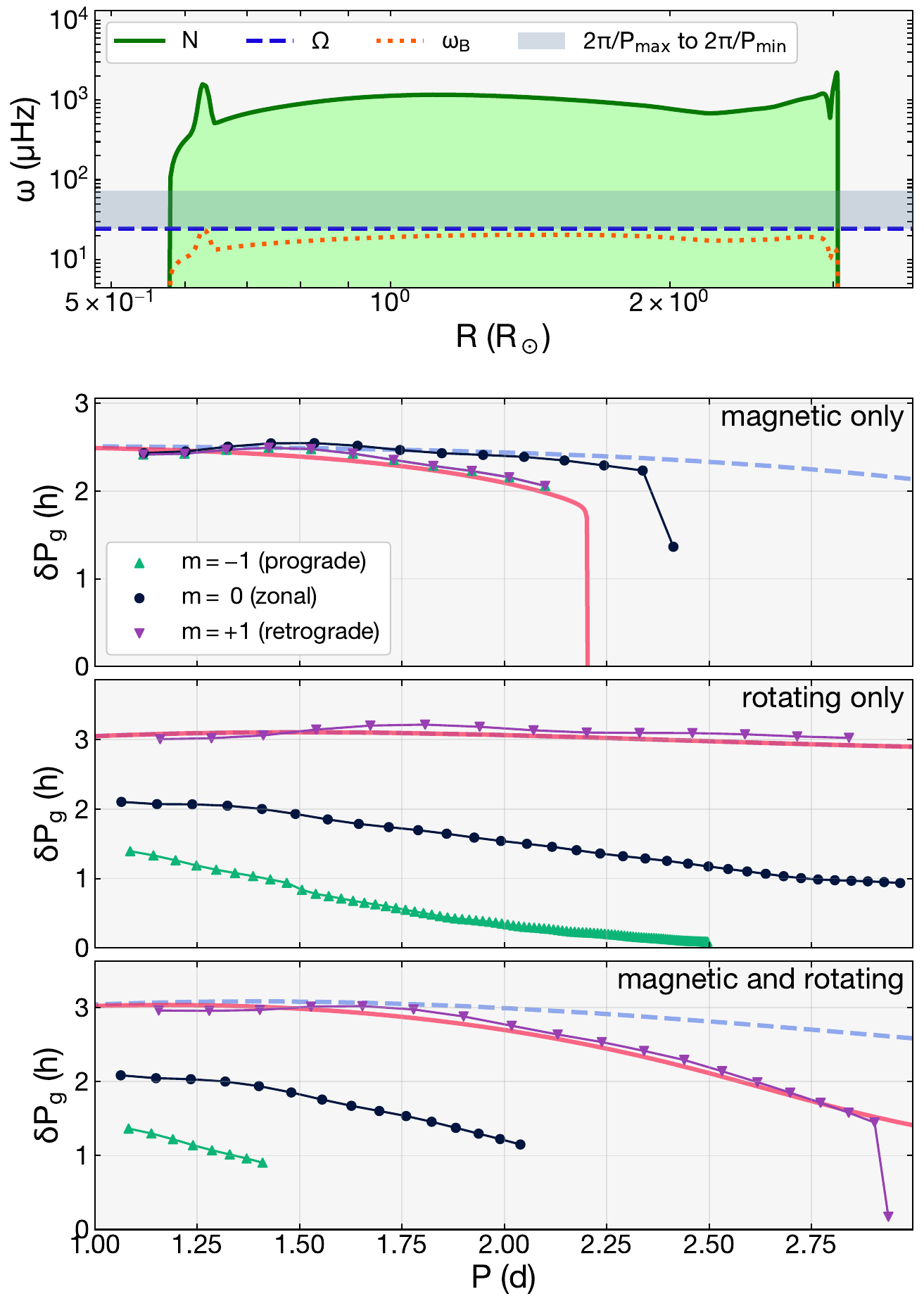}
    \caption{Same as \autoref{fig:dipole_modes_MS_0}, but for a young SPB analogue (MS-6.0-young).}
    \label{fig:dipole_modes_MS_2}
\end{figure}

\begin{figure}
    \centering
    \includegraphics[width=0.46\textwidth]{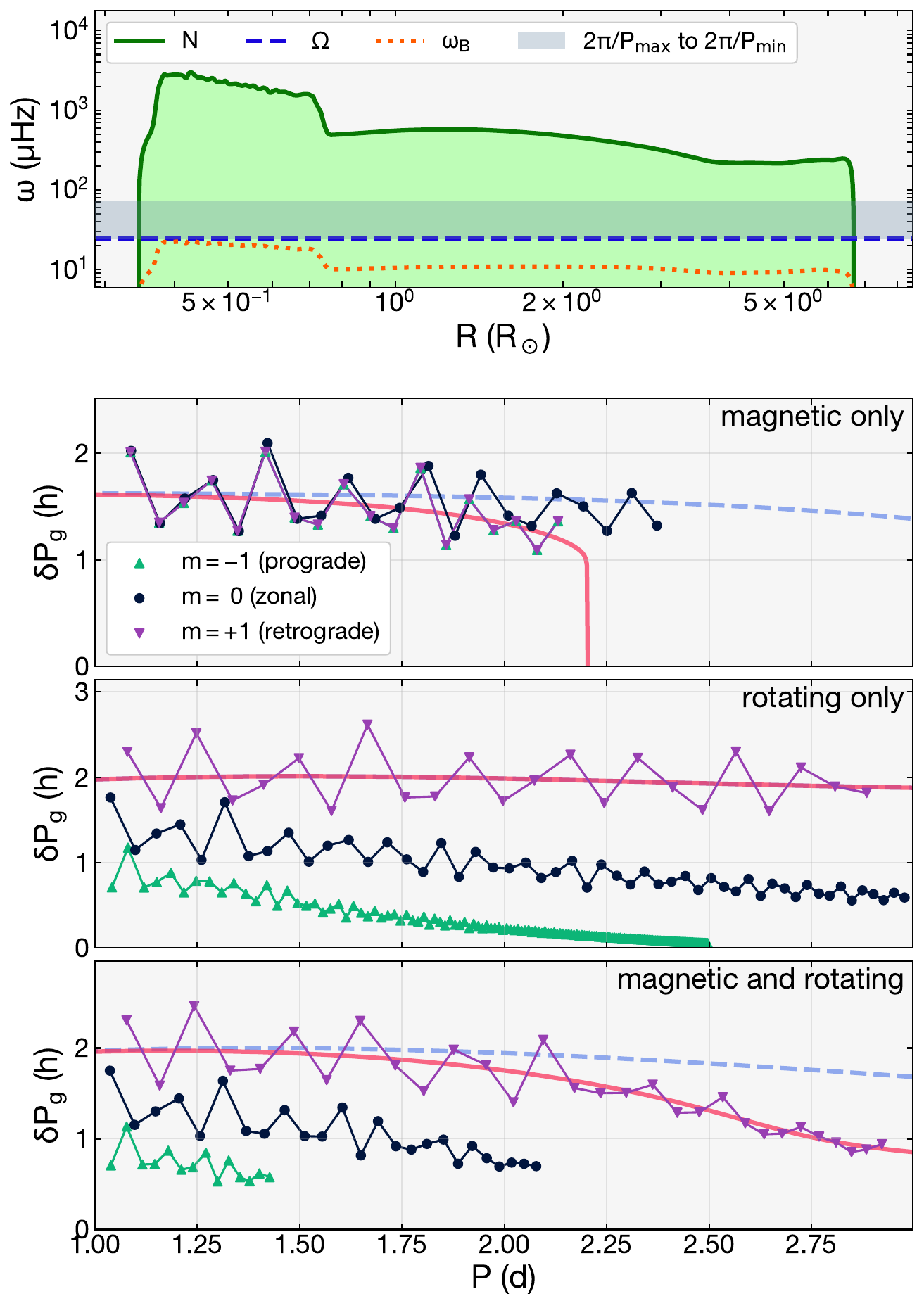}
    \caption{Same as \autoref{fig:dipole_modes_MS_0}, but for a evolved SPB analogue (MS-6.0-evolved).}
    \label{fig:dipole_modes_MS_3}
\end{figure}

Stars with masses $\gtrsim\!\!1.3M_\odot$ have radiative envelopes and convective cores on the main sequence.
Therefore, such stars may pulsate in g modes which are directly detectable, without needing to be disentangled from p modes as in solar-like oscillators.
Such oscillators are ubiquitous: as discussed previously, they include $\gamma$ Dor (AF-type) and SPB (B-type) variables.
The pulsations are driven by coherent mechanisms such as convective flux blocking \citep[in $\gamma$ Dors;][]{guzik2002proposed,dupret2004theoretical} and the $\kappa$ mechanism \citep[in B-type pulsators;][]{gautschy1993non,dziembowski1993opacity}.
This is in contrast to the broadband, stochastic driving present in solar-like oscillators \citep{samadi2015stellar}.
Crucially, in these pulsators, there is no guarantee that the measurable modes are complete over some observed frequency range.
The selection mechanism for mode excitation is poorly understood, and the asteroseismic power spectra are often sparse.
Observational studies of such pulsators thus typically apply a forward-modeling approach based on the identified modes \citep[e.g.,][]{aerts2018forward}, which rely on good models for predicting observed oscillation spectra.

In this Section, we primarily focus on the period spacing pattern $\delta P_g=\delta P_g(P)$ for modes of a given $m$.
This is a standard observable in the study of main-sequence pulsators.
The period spacing pattern is known to encode the rotation rate of the star \citep[through an overall slope;][]{bouabid2013effects,ouazzani2016new}, as well as the presence of buoyancy glitches \citep[e.g.,][]{miglio2008probing}.

We first calculate the dipole oscillation modes for two $\gamma$ Dor-like models, one near the zero-age main sequence (MS-1.5-young) and one near the terminal-age main sequence (MS-1.5-evolved), shown respectively in Figures \ref{fig:dipole_modes_MS_0} and \ref{fig:dipole_modes_MS_1}.
The chief difference between these models is that the convective core in the latter model has had time to develop a large compositional gradient at the base of its radiative envelope: this produces a jump in $N$ (see the top panel of \autoref{fig:dipole_modes_MS_1}).
Qualitatively, this sharp feature in $N$ results in a trapping phenomenon which results in a period spacing $\delta P_g$ which oscillates as a function of mode period $P$ \citep{miglio2008probing,pedersen2018shape,vanlaer2023feasibility}.
We adopt a fairly typical core rotation rate of $1.5\,\mathrm{d}$ to accentuate the effects of rotation \citep{van2016interior,li2020gravity}.
Unlike in the red giant model described in Section \ref{redgiant} (where realistic rotation rates are small, such that $q\!\ll\!1$), rotation in the MS models is fast enough to cause frequency splittings/shifts which are nonlinear with respect to $\Omega$.

The lower panels of Figures \ref{fig:dipole_modes_MS_0} and \ref{fig:dipole_modes_MS_1} show $\delta P_g$ versus $P_g$ for the young and evolved $1.5M_\odot$ models, under the effects of magnetism and rotation individually as well as simultaneously.
First, since rotation distinguishes between prograde and retrograde modes, the slope it imparts onto the period spacing pattern is different for the $m\!=\!+1$ and $m\!=\!-1$ modes.
In contrast, the oscillation modes are not sensitive to the overall sign of the magnetic field, and thus magnetism affects the $m\!=\!\pm1$ modes identically (but still differently than the $m\!=\!0$ mode).

Moreover, while rotation produces values of $P_g$ which vary fairly linearly with $P$, magnetism produces a curvature in the pattern, especially near suppression.
This effect is similar to what was demonstrated by \citet{dhouib2022detecting} in the case of a purely toroidal field.
In particular, when the maximum allowed value of $a=a_{\mathrm{crit}}$ is determined by connection to the evanescent region (rather than the presence of a critical Alfv\'en latitude), the asymptotic expression (in \autoref{asymptotic}) predicts that $\delta P_g$ sharply approaches zero at $\omega\approx\omega_B$.
This is because the term $\propto\mathrm{d}\ln\lambda/\mathrm{d}\ln a$ in \autoref{asymptotic} diverges at radii where the main magnetogravity wave branch connects to the slow branch described by \citet{lecoanet2017conversion} and \citet{rui2023gravity}.
In reality, there is not likely to be an infinitely dense forest of modes, since the asymptotic formula is based on a linear approximation which is likely to break down close to suppression.
Nevertheless, the curvature is conspicuous, especially for the young model, where the period spacing drops from its high-frequency value by $\simeq\!50\%$ near the critical frequency.
Moreover, this curvature is apparent even when rotation is included alongside magnetism, with the added feature that fast rotation can cause the $m\!=\!+1$ and $m\!=\!-1$ modes to become magnetically suppressed at very different frequencies.
\edit{This curvature effect on the period spacing pattern is very different than those caused by inertial-mode coupling in main-sequence convective cores \citep[which manifest as isolated ``dips'';][]{tokuno2022asteroseismology} and mode-trapping near strong compositional gradients outside of those cores \citep[which manifests as ``oscillations'';][]{miglio2008probing}.}

This sharp curvature feature is not adequately captured by any low-order perturbative treatment of magnetism.
To make comparison to the perturbative prediction generous, we expand \autoref{asymptotic} around $a=0$, while treating rotation non-perturbatively \citep[through the traditional approximation of rotation, cf.][]{van2020detecting}.
The effect of magnetism then enters the period spacing earliest through $a^2\propto\omega_B^4/\omega^4\propto B^2/B_{\mathrm{crit}}^2$ \citep[as predicted by][]{cantiello2016asteroseismic}.
Specifically, defining $\lambda_H$ to be the eigenvalue calculated including rotation only, we have
\begin{equation} \label{pertmagonly}
    \begin{split}
        \delta\bar{P}_g &\approx \frac{2\pi^2}{\sqrt{\lambda_H}}\left(\int\left(1 + \frac{1}{2}\frac{\mathrm{d}\ln\lambda_H}{\mathrm{d}\ln q}\right)\frac{N}{r}\,\mathrm{d}r\right)^{-1} \\
        &- \frac{16\pi^6\lambda'}{\lambda_H^{3/2}}\bar{P}^{-4}\left(\int\omega_B^4\left(1 + \frac{1}{2}\frac{\mathrm{d}\ln\lambda_H}{\mathrm{d}\ln q}\right)\frac{N}{r}\,\mathrm{d}r\right)^{-1}\mathrm{,}
    \end{split}
\end{equation}

\noindent where we have used
\begin{equation}
    \left(\frac{\mathrm{d}\ln\lambda}{\mathrm{d}\ln a}\right)_{a=0} = 0\mathrm{.}
\end{equation}

In addition to lacking the suppression phenomenon entirely, the perturbative prediction (shown for the $m\!=\!+1$ mode as the blue-dashed lines in Figures \ref{fig:dipole_modes_MS_0} and \ref{fig:dipole_modes_MS_1}) dramatically underestimates the magnetic curvature predicted by the full TARM-based formalism.
To further demonstrate this point, in \autoref{fig:parameter_space}, we show contours where the perturbative estimate misestimates the integrand of the integral in the asymptotic formula (\autoref{asymptotic}) by $10\%$ and $50\%$, respectively.
As expected, departure from the full TARM calculation becomes increasingly severe close to suppression.
Non-perturbative effects must therefore be taken into account predicting the frequency spectrum close to $\omega_{\mathrm{crit}}$.
For example, the magnetic ``sawtooth'' pattern in the period spacing pattern predicted by some authors \citep{prat2019period,prat2020period,van2020detecting} was derived using perturbation theory at low frequencies, and preliminary results suggest that this feature does not appear once magnetism is incorporated non-perturbatively (Dhouib et al., in prep.).

An important observation is that the magnetically induced curvature in the period spacing pattern is more conspicuous in the young model than in the evolved one.
This is because the relative magnetic frequency shifts are primarily determined by the quantity $\langle\omega_B^4\rangle_g^{1/4}/\bar{\omega}_0$ (as shown in Section \ref{redgiant}), which is maximized when as many layers of the star have $\omega_B\sim\bar{\omega}_0$ as possible.
However, within our physical picture, the entire oscillation mode becomes suppressed when even a small layer of the star has $\bar{\omega}_0\lesssim\omega_{\mathrm{crit}}\sim\omega_B$.
Because $N$ accounts for most of the variation of $\omega_B\propto\sqrt{N}$ (the Prendergast field we adopt varies comparatively more slowly with radius), $\omega_B$ is a much broader function of $r$ in the young model versus in the evolved one, where it is peaked at the composition gradient at the lower boundary of the radiative envelope.
Therefore, the young model reaches a larger maximum value of $\langle\omega_B^4\rangle_g^{1/4}/\bar{\omega}_0$ than the evolved one, and furthermore in general attains large values of $\langle\omega_B^4\rangle_g^{1/4}/\bar{\omega}_0$ over a wider frequency range.
This heuristic explanation is even stronger for higher-order terms in the perturbative expansion, which involve buoyant integrals of higher powers of $\omega_B^4/\bar{\omega}_0^4$.

The magnetic curvature is in principle detectable even in evolved main-sequence pulsators, as long as it can be deconvolved from other effects.
It should be noted that the typical uncertainties in $\gamma$ Dor period spacings in \textit{Kepler} are small, comparable to the marker sizes of Figures \ref{fig:dipole_modes_MS_0} and \ref{fig:dipole_modes_MS_1} \citep{vanreeth2015detecting,li2020gravity}.
Moreover, because of the sensitivity of the magnetic curvature in the period spacing pattern to the compositional profile, strongly magnetized main-sequence pulsators may be a promising avenue for constraining mixing processes.
However, in nonasymptotic cases where sharp features in the buoyancy profile are expected, the limitations of the TARM must carefully be considered.

For completeness, we also examine young (MS-6.0-young) and evolved (MS-6.0-evolved) SPB analogues, with masses $6M_\odot$ (Figures \ref{fig:dipole_modes_MS_2} and \ref{fig:dipole_modes_MS_3}, respectively).
The qualitative features of the period spacing pattern are similar, except that the peak in $\omega_B$ at the base of the radiative region in the young model (due to the peak in $N$) exceeds the value of $\omega_B$ throughout the rest of the cavity.
Therefore, for similar reasons as in the evolved $\gamma$ Dor model, the curvature in the period spacing pattern due to magnetism is not as prominent as in the young $\gamma$ Dor model.

\begin{figure}
    \centering
    \includegraphics[width=0.49\textwidth]{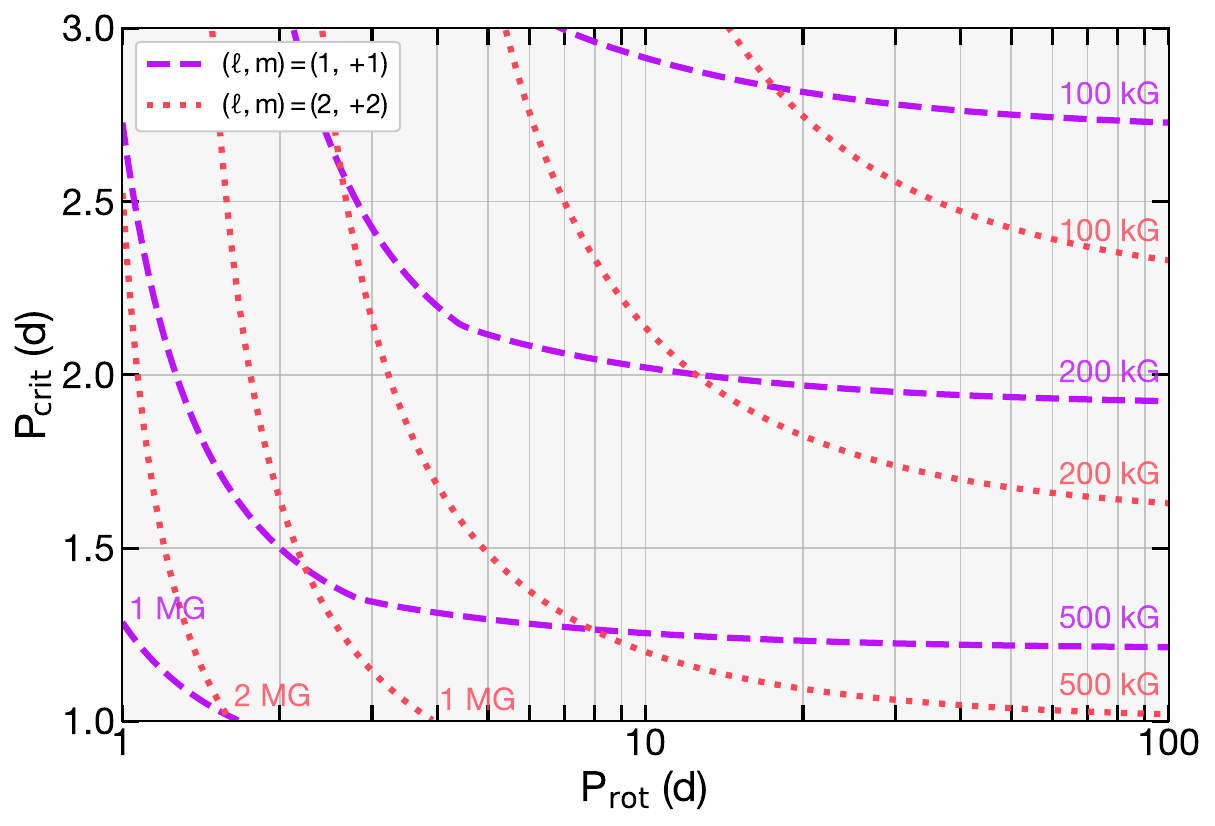}
    \caption{The critical period $P_{\mathrm{crit}}$ against the rotation period $P_{\mathrm{rot}}$ for a young SPB-like model (MS-6.0-young), for fixed values of the field near the compositional gradient at the base of the radiative envelope (which most easily experiences magnetic suppression).
    $P_{\mathrm{crit}}$ is given in the inertial frame.}
    \label{fig:suppressed_diagram}
\end{figure}

For illustrative purposes, we calculate the critical mode frequencies for a variety of internal magnetic fields and rotation rates, using the MS-6.0-young model.
\autoref{fig:suppressed_diagram} shows the critical mode period $P_{\mathrm{crit}}=2\pi/\omega_{\mathrm{crit}}$ for the dipole and quadrupole prograde sectoral modes.
Interestingly, although rotation is expected to make prograde modes suppress at higher frequencies in the co-rotating frame (see \autoref{fig:parameter_space}), higher rotation rates actually cause modes to suppress at \textit{lower} frequencies in the inertial frame.
Simultaneous knowledge of the suppression frequency for one identified mode branch together with the rotation rate should be sufficient to make a model-dependent estimate of the magnetic field at the interior of the star.
Alternatively, while potentially challenging, simultaneous measurement of the suppression frequencies for two identified mode branches may be able to put a constraint on both the internal magnetic field as well as rotation rate.
Because the shapes of the contours in \autoref{fig:suppressed_diagram} are largely determined by change-of-frame effects (vis-\`a-vis \autoref{converttoinertial}), the latter method is most viable when the two mode branches have different azimuthal order $m$.

Roughly $\sim\!\!10\%$ of massive dwarf stars possess significant (inclined dipolar) fossil fields up to tens of kilogauss at their surfaces \citep{grunhut2016mimes,shultz2019magnetic}.
Such fields may be strong enough in the interiors of such stars to suppress low-frequency g-mode oscillations.
Recently, \citet{lecoanet2022asteroseismic} attributed missing low-frequency modes in the magnetic SPB star HD 43317 \citep[observed with CoRoT;][]{buysschaert2017magnetic,buysschaert2018forward} to magnetic suppression caused by a near-core radial field $B_r\!\simeq\!500\,\mathrm{kG}$.
As in our MS-6.0-early model, suppression in their model occurs when $\omega_{\mathrm{crit}}>\omega$ in the compositional peak in $N$ at the base of the radiative cavity (see their Figure 2).
Moreover, \citet{aerts2017interior} predict that core dynamos in B-type (AF-type) pulsators may produce strong magnetic fields $20$--$400\,\mathrm{kG}$ ($0.1$--$3\mathrm{kG}$) where non-perturbative magnetic effects may be realized.
Magnetic g-mode main-sequence stars thus appear to be natural environments to observe g modes which are non-perturbatively modified by magnetism.

Pulsators in the $\gamma$ Dor mass range may also possess influential magnetic fields \citep{aerts2021rossby}.
Surface fields of hundreds to thousands of gauss are typical of the enigmatic family of rapidly oscillating Ap-type (roAp) stars \citep{hubrig2004new}, and the magnetic field is believed to play an important role in the (still not fully understood) driving mechanism of their high-overtone p-mode oscillations \citep{gautschy1998drive,balmforth2001excitation}.
It has been speculated \citep[e.g., by][]{handler2011hybrid} and claimed \citep{balona2011kepler} that some roAp stars may also pulsate in g modes (on the basis of overlap between roAp and $\gamma$ Dor stars on the Hertzsprung--Russell diagram).
However, this is far from certain.
On the basis of non-adiabatic mode calculations, \citet{murphy2020first} argue that high-order g modes are likely to be very efficiently damped, possibly explaining the current lack of observed hybrid $\gamma$ Dor/roAp pulsators.
However, if roAp stars containing high-order g modes do turn out to exist, they would serve as ideal laboratories for strong magnetogravity waves.
Moreover, the understanding of high-order magnetogravity waves presented in this work may extend some insight into the behavior of low-order magnetic g modes (for which the asymptotic limit is not appropriate).

\subsection{Future prospects}

This work presents a non-perturbative formalism for calculating the g-mode oscillation frequencies of a magnetized and rotating star, including both effects asymptotically (i.e., applying the TARM).
We have considered only with the case where the magnetic field is dipolar and aligned with the rotational axis.
As test examples we have only applied it to red giant cores and g-mode pulsators on the main sequence.
Here, we describe future possible directions of study in relation to the TARM formalism, and potential extensions.

This work represents a joint generalization of the traditional approximation of rotation \citep{lee1997low} and an analogous approximation for a purely dipolar magnetic field \citep{rui2023gravity}, in order to non-perturbatively incorporate the effects of both.
Generalizations of the traditional approximation have, in the past, also incorporated centrifugal distortion \citep{mathis2019traditional,dhouib2021traditional,dhouib2021traditional2}, differential rotation \citep{ogilvie2004tidal,mathis2009transport,vanreeth2018sensitivity,dhouib2021traditional2}, and axisymmetric toroidal fields, both with constant Alfv\'en and rotation frequencies \citep{mathis2011low} as well as with more general field geometries together with differential rotation \citep{dhouib2022detecting}.
Based on observational demands (or theoretical intrigue), it is likely possible to add any combination of these effects to the operator $\mathcal{L}^{m,b,q}$ defined in \autoref{L}.
Although $\lambda$ would then be a function of more than two dimensionless parameters, such an approach would retain much of the advantage of non-perturbatively capturing complex rotational/magnetic effects while only interpolating over a precomputed eigenvalue grid.

Unlike \citet{rui2023gravity}, this work has focused on the regime where suppression is not likely to occur, i.e., when there are no Alfv\'en resonances on the domain and where the slow magnetic branch has been ignored.
We have ignored modes with these effects because their observational implications are unclear, but the behavior of the operator $\mathcal{L}^{m,b,q}$ in this regime is an extremely rich mathematical problem with so far unexplored structure.
\edit{\citet{rui2023gravity} find that solutions with $b\!>\!1$ develop sharp fluid features at the Alfv\'en-resonant critical latitudes, where processes such as phase-mixing are likely to efficiently damp the waves.
In this regime, the magnetic operator in \autoref{mto} is of Boyd-type \citep{boyd1981sturm}, and the interior singularities give dissipation an important role in determining the physically appropriate branch cut.
The eigenvalues $\lambda$ for the $b\!>\!1$ are thus not guaranteed to be real even in the formal limit where dissipation is taken to zero (and the numerical results of \citet{rui2023gravity} suggest that they are not).}
For reasons of scope, we have also ignored magneto-Rossby waves and magnetically stabilized gravity waves \citep{rui2023gravity}, which do not connect to any spherical harmonic in the limits $a,q\rightarrow0$.
These, too, may conceal detectable predictions which are implied by the breakdown of positive-(semi)definiteness of $\mathcal{L}^{m,b,q}$.

\edit{As such, our calculations also do not capture the coupling between magnetic g modes and magneto-inertial modes which propagate in the convective core of intermediate-mass main-sequence stars \citep[within which dynamo-generated magnetic fields are expected;][]{Brunetal2005,Featherstoneetal2009}.
Coupling with inertial modes is known to result in isolated dips in the $\delta P_g$--$P$ diagram at frequencies corresponding to those of inertial modes.
This effect provides a seismic probe of the core rotation rates of such stars \citep{Ouazzanietal2020,Saioetal2021,tokuno2022asteroseismology}.
In the future, it may be interesting to explore how this picture is modified by magnetism, and whether similar inference of the magnetic field in these convective cores is possible.
We emphasize that coupling to (magneto-)inertial waves produces \textit{localized} dip features in the period spacing pattern, and is very different than the \textit{global} curvature in the pattern predicted by this work.}

While we have only explicitly modeled analogues of $\gamma$ Dor and SPB stars, our analysis applies to any magnetized pulsator with pulsations of high radial order.
This includes compact pulsators such as white dwarfs and hot subdwarfs.
Since both of these species result from red giants whose envelopes have been lost (either in isolation or through binary evolution), it is natural to expect that they will retain the strong fields believed to cause dipole suppression in red giants.
While a small handful of magnetized hot subdwarfs ($100$s of kG) are known \citep{pelisoli2022discovery}, white dwarfs with kilogauss surface fields are believed to make up a fourth of all white dwarfs \citep{cuadrado2004discovery,valyavin2006search}, and a number of magnetized white dwarfs with fields up to hundreds of megagauss have been discovered \citep{kepler2013magnetic,bagnulo2021new}.
The latter fields are likely to be so strong that they outright suppress g mode oscillations altogether \citep{lecoanet2017conversion}.
However, it may be possible for a white dwarf to have a field strong enough to significantly shift the frequencies of the g modes, without being not strong enough to suppress them outright.

While a dipolar field is expected at the surfaces of stars with fossil fields \citep{braithwaite2006stable,duez2010relaxed}, that field need not be aligned with the rotation axis \citep{duez2011numerical,keszthelyi2023magnetism}, and is unlikely to be dipolar at all if the field is generated by a dynamo.
In the perturbative regime, \citet{mathis2023asymmetries} recently characterized the frequency shifts associated to an inclined dipole field.
Extending the TARM formalism to describe a non-axisymmetric horizontal field dependence requires solving for the eigenvalues of families of two-dimensional differential operators over the sphere, rather than a one-dimensional one (as in $\mathcal{L}^{m,b,q}$), and this analysis would need to be repeated for every different horizontal field dependence desired.
Nevertheless, near suppression, departures in the frequency shifts from the perturbative theory are likely, and may be required for accurate magnetic field inference in this regime.

Finally, low-frequency propagating gravity waves are one of the best candidates for the strong angular momentum transport needed in stellar radiative zones to reproduce the observed internal rotation revealed in all types of stars by helio- and asteroseismology \citep[e.g.,][]{schatzman1993transport,zahn1997rotation,charbonnel2005influence,aerts2015age,rogers2015differential,pinccon2017can,neiner2020transport}.
The manner in which this wave-mediated angular momentum transport occurs can be significantly modified by the presence of a magnetic field.
In general, the net angular momentum flux implied by this mechanism is given by the sum of the wave-induced Reynolds and Maxwell contributions to the stress tensor.
The relevant gravity waves are precisely those which are most strongly affected by the combined action of rotation and magnetism \citep[see, e.g.,][in the case of weak, shellular differential rotation and a purely toroidal field with constant Alfv\'en frequency]{mathis2012low2}.
Because our TARM-based formalism is relevant to exactly this kind of wave, its application to this problem is likely to yield insights into the rotational state and internal chemical mixing of rotating, magnetic stars.

\section{Conclusion} \label{conclude}

Rapidly evolving progress in observational magnetoasteroseismology demands refinements in our theoretical understanding of magnetic effects on stellar pulsations.
In this work, we develop a formalism for incorporating the effects of an aligned dipole magnetic field into g mode calculations, valid for rapidly rotating stars.
This method relies on an asymptotic treatment of magnetism and rotation (under a ``traditional approximation of rotation and magnetism''), and can be partitioned into two main steps:
\begin{enumerate}
    \item Calculate the eigenvalues $\lambda$ of the horizontal differential operator $\mathcal{L}^{m,b,q}$ (\autoref{L}) as a function of the dimensionless magnetic and rotational parameters $a=\omega_B^2/\bar{\omega}^2$ and $q=2\Omega/\bar{\omega}$.
    
    \item In either an asymptotic mode formula (\autoref{ugly}) or a non-asymptotic numerical scheme (e.g., shooting; Section \ref{nonasymptotic}), include the effects of magnetism and rotation by replacing $\ell(\ell+1)$ throughout the star with a suitably interpolated $\lambda$, calculated using the magnetic and rotational profiles.
\end{enumerate}

These steps are done relatively independently of each other: once the eigenvalues $\lambda$ are computed once over a sufficiently large grid of $a$ and $q$ (for the desired $\ell$ and $m$), they do not need to be calculated again for any individual stellar model.
Moreover, modifications to existing mode solving procedures are ``minimal'' in the sense of being localized to the interpolation of $\lambda$ and its substitution into the relevant equations.

As proofs of concept, we have computed the g modes in the cores of red giants as well as in the radiative envelopes of high-mass main-sequence stars.
In both cases, strong magnetic fields tend to decrease the period spacing significantly more than is suggested by the perturbative theory, especially for low frequencies close to the critical frequency $\omega_{\mathrm{crit}}\sim\sqrt{Nv_{Ar}/r}$.
This results in a curvature in the period spacing pattern which can in some cases be very conspicuous (e.g., \autoref{fig:dipole_modes_MS_0}).
Non-perturbative effects may also introduce asymmetry in the dipole frequency shifts which is not predicted by perturbation theory.

This regime is expected to be directly realized in the SPB star described by \citet{lecoanet2022asteroseismic} and some of the red giants described by \citet{deheuvels2023strong}.
Refined understanding of these effects is therefore prerequisite to perform accurate magnetic field inference using asteroseismology.

\section*{Acknowledgements}
We graciously thank Conny Aerts and Jim Fuller for their very helpful and thorough comments.
N.Z.R. acknowledges support from the National Science Foundation Graduate Research Fellowship under Grant No. DGE‐1745301.
J.M.J.O. acknowledges support from NASA through the NASA Hubble Fellowship grant HST-HF2-51517.001-A, awarded by STScI.
STScI is operated by the Association of Universities for Research in Astronomy, Incorporated, under NASA contract NAS5-26555.
S.M. acknowledges support from the European Research Council through HORIZON ERC SyG Grant 4D-STAR 101071505, from the CNES SOHO-GOLF and PLATO grants at CEA-DAp, and from PNPS (CNRS/INSU).
\edit{We thank the referee for their positive and constructive comments which improved the quality of the final manuscript.}

\section*{Data Availability}

An accompanying Zenodo repository\footnote{\href{https://zenodo.org/record/8329841}{https://zenodo.org/record/8329841}} includes the following resources:
\begin{itemize}
    \item Eigenvalues $\lambda$ of $\mathcal{L}^{m,b,q}$ as a function of $a$ and $q$, computed for all dipole ($\ell=1$) and quadrupole ($\ell=2$) modes.
    \item MESA inlists and stellar profiles (in GYRE format) for the stellar models used in this work.
\end{itemize}



\bibliographystyle{mnras}
\bibliography{bibliography} 




\appendix

\bsp	
\label{lastpage}
\end{document}